\documentclass[11pt]{article}
\usepackage{amsfonts}
\usepackage{amssymb}
\def\whitebox{{\hbox{\hskip 1pt
        \vrule height 6pt depth 1.5pt
        \lower 1.5pt\vbox to 7.5pt{\hrule width
                  3.2pt\vfill\hrule width 3.2pt}%
        \vrule height 6pt depth 1.5pt
        \hskip 1pt } }}
\def\qed{\ifhmode\allowbreak\else\nobreak\fi\hfill\quad\nobreak
     \whitebox\medbreak}

\newcommand{\C}{{\cal C}}

\newcommand{\cc}{{\bf c}}
\newcommand{\remove}[1]{}

\newcommand{\A}{{\cal  A}}
\newcommand{\M}{{\cal  M}}

\newcommand{\y}{{\tilde y}}
\newcommand{\yb}{{\hat y}}

\newcommand{\yy}{{z}}

\newcommand{\G}{{\cal G}}

\newcommand{\F}{{\cal F}}

\newcommand{\dd}{{\bar d}}
\newcommand{\proof}{\noindent{\bf Proof. }}

\newtheorem{lemma}{Lemma}
\newtheorem{theorem}{Theorem}

\newtheorem{corollary}{Corollary}
\newtheorem{definition}{Definition}

\begin{document}
\title{Constraining the Number of Positive Responses in Adaptive, Non-Adaptive, and Two-Stage Group Testing}
\author{\sc Annalisa De Bonis\\
Diparitmento di Informatica, Universit\`a di Salerno,\\ Fisciano (SA), Italy.\\
email: \texttt{debonis@dia.unisa.it}}

\date{}

\maketitle
\begin{abstract}
{\em Group testing} is  a well known search problem that consists in detecting the defective members of a set of objects $O$ by performing tests on properly chosen subsets ({\em pools}) of the given set $O$. In classical group testing the goal is to find all defectives by using as few tests as possible. 
We consider a variant of classical group testing in  which  one is concerned not only with minimizing the total number of tests but aims also at reducing the number of tests involving defective elements.   The rationale behind this search model is that in many practical applications the devices used for the tests are subject to deterioration due to exposure to or interaction with the defective elements.    
 In this paper we consider   adaptive, non-adaptive and two-stage group testing.  For all three considered scenarios, we derive
   upper and lower bounds on the number of ``yes" responses that must be admitted by any strategy  performing at most a certain number $t$ of tests. In particular, for the adaptive case we provide an algorithm that  uses a number of ``yes" responses that exceeds the given lower bound by a small constant.  Interestingly, this bound can be asymptotically attained also by our two-stage algorithm, which is a phenomenon analogous to the one occurring in classical group testing. For the non-adaptive scenario we give almost matching upper and lower bounds on the number of ``yes" responses. In particular, we give two constructions both achieving the same asymptotic bound. An interesting feature of one of these constructions is that it is an explicit construction. The bounds for the non-adaptive  and the two-stage cases follow from the  bounds on the optimal sizes of new variants of $d$-cover free families and $(p,d)$-cover free families introduced in this paper, which we believe may be of interest also in other contexts.  
\end{abstract}

 \section{Introduction}

 {\em Group testing} is  a well known search paradigm that consists in detecting the defective members  of a set of objects $O$ by performing tests on properly chosen subsets ({\em pools}) of the given set $O$.  A test yields a ``yes" response if the tested pool contains one or more defective elements, and a ``no" response otherwise. The goal is to find all defectives by using as few tests as possible. Group testing origins date back to World War II when it was introduced as a possible technique for mass blood testing  \cite{dor}.  Since then group testing has found applications in a wide variety of situations ranging from conflict resolution algorithms for multiple-access systems \cite{TCS}, \cite{wolf},  fault diagnosis in optical networks \cite{optical}, quality control in product testing \cite{sobel}, failure  detection in wireless sensor networks \cite{faultSensor}, data compression \cite{hl}, and many others. 
Among the modern applications of group testing,  some of the most important are related to the field of  molecular biology, where group testing is especially employed in the design of  screening experiments. Du and Hwang  \cite{dh1}  provide an extensive coverage of the most relevant applications of group testing  in this area. 

The different contexts to which group testing applies often call for variations of the classical model that best adapt to the characteristics of the problems. These variants concern the test model   \cite{jcb}, \cite{tcs3}, \cite{iplPeter},  \cite{TCS},  \cite{it}, the number of pursued defective elements  \cite{majority}, \cite{disjointGT}, as well as the structure of the test groups \cite{interval}, \cite{disjointGT},   \cite{tcs1}. 

In this paper,
we consider a variant of the classical model in  which  one is concerned not only with minimizing the total number of tests but aims also at reducing the number of tests involving defective elements.  Therefore, the test groups should be structured so as to reduce the number of groups intersecting the set of defectives.  The rationale behind this search model is that in many practical applications the devices used for the tests are subject to deterioration due to exposure to or interaction with the defective elements. In some contexts, the positive groups may even represent a risk for 
the safety of the persons that perform the tests.  An example of such applications are leak testing procedures aimed at guaranteeing the safety of sealed radioactive sources  \cite{radio1,radio2}. Radioactive sources are widely used in medical, industrial and agricultural applications, as well as in scientific research.  
Sealed sources are small metal containers in which radioactive material is sealed. 
 As long as the sealed sources are handled correctly and  the enclosing capsules are intact, they do not represent a health hazard.  According  to  
 the radiation safety standards, sealed radioactive sources should be tested at regular intervals in order to verify  the integrity of the capsules. Leak testing procedures are crucial in preventing contamination of facilities and personnel due to the escape of radioactive material. However, these procedures put the safety personnel at the risk of being exposed to radiation whenever a leak in the tested sources is present. 
Commonly, when not used, the sources are stored in lead-shielded drawers. In order to be tested for leakage, sources are removed one at time from the storage area and wiped with absorbent paper or  a cotton swab held by a long pair of forceps.   The wipe sample is then analyzed for radioactive contamination.
An alternative procedure consists in testing the sources in groups. To this aim, the sources are not removed from the shielded storage  drawer and a wipe sample is taken from the upper surface of the storage drawer. If the sample is contaminated then at least one source in the tested storage drawer is leaking;  otherwise all sources in the drawer  are intact. This idea suggests the use of group testing in leak testing procedures.
Since leak testing procedures  expose to risk the personnel that perform the tests on contaminated wipe samples,
the number of positive tests admitted by the group testing procedure should depend on the dose of radiation which is judged to be of no danger for the  health. Obviously, the total number of tests should also be taken into account in order to reduce the costs and the work load of the safety personnel.     Trivially, the  procedure that tests all elements individually attains the minimum number of positive responses, which is equal to the number of defectives in the input set. While this procedure may be an option when the danger implied by testing positive samples is extremely high, many practical applications call for procedures that can be tuned to obtain the desired tradeoff between the number of ``admissible" positive responses and the total number of tests.

\subsection{Summary of results}
 We consider   {\em adaptive}, {\em non-adaptive} and {\em two-stage} group testing procedures. In adaptive group testing, at each step the algorithm decides which group to test  by observing the responses of the  previous tests. 
For classical group testing, there exist adaptive strategies that  achieve the information theoretic lower bound $\Omega(d\log(n/d))$, where $n$ is the total number of elements and $d$ is the upper bound on the number of defectives. We will prove that in our model
 any adaptive algorithm must tolerate a number $y$ of positive responses of order $\Omega\left(\frac{d\log(n/d)}{\log (et/y)}\right)$, where $t$ is the total number of tests (i.e., the total of positive and negative tests), and give an adaptive algorithm that attains this lower bound. In fact, the exact values of the two bounds differ 
 by a little constant. Therefore, if we require that $y=O(t^{1-c})$, for any positive constant $c<1$, then the number of positive responses admitted by our optimal algorithm  grows as  $O\left(\frac{d\log(n/d))}{\log t}\right)$. 
 
In many practical scenarios adaptive strategies are useless  due to the fact that assembling the groups for the tests may be very time consuming and that some kind of group tests may take long time to give a response.
 In such applications, it is preferable to use non-adaptive strategies, i.e., strategies in which all tests are decided in advance and can be 
 performed in parallel. 
 Non-adaptive group testing strategies are  much more costly than adaptive algorithms. Indeed,  the  minimum number of tests used by these procedures is equal to the minimum length of certain combinatorial structures known under the name of $d$-{\em superimposed codes}
  (or equivalently, the minimum size of the ground set of $d$-{\em cover free families }) \cite {dyry}, \cite{erff}, \cite{ks}.
The known bounds for these combinatorial structures imply that the number of tests of any non-adaptive group testing algorithm is lower bounded by
$\Omega((d^2/ \log d) \log n)$ and that there exist non-adaptive group testing algorithms that use $O(d^2\log n)$ tests. 
 In order to study the non-adaptive case under our model, we will introduce a new variant of $d$-cover free families and derive upper and lower bounds on the size of these combinatorial structures. In particular, we will 
 show that any non-adaptive algorithm for our group testing problem must admit a number of positive responses $y$ of order
 $\Omega\left(\frac{d^2} {\log \left(\frac {etd^2}y\right)}\log n\right)$ and give 
 two almost optimal algorithms that must tolerate a number of positive responses $y$ of order 
 $O\left(\frac{d^2} {\log\left( \frac {et}y\right)}\log n\right)$.  An inte\-resting feature of one of these constructions consists in being an {\em explicit} construction, in that there exists an efficient algorithm to design the underlying combinatorial structure. Interestingly, the gap between the above upper and lower bounds decreases as the ratio between the total number $t$ of tests   and the number $y$ of positive responses admitted by the algorithm increases.
Fot $y =t$, i.e., for algorithms that admit an unlimited number of positive responses, this  gap  is equal to that 
 existing between the best upper and lower  bounds on the minimum number of tests for classical group testing.   Closing this gap is considered
  a major open problem in extremal combinatorics.

 In \cite{dgv} it has been proved that by allowing a little adaptiveness it is possible to dramatically improve on the number of tests
 used by non-adaptive group testing procedures.
 Indeed, the authors of that paper gave a trivial two-stage algorithm for classical
 group testing that uses the same number of tests of the best adaptive procedures. A trivial two-stage group testing algorithm consists
 of two  non-adaptive stages. In the first stage the algorithm performs parallel tests  on certain pools of elements with the aim of selecting
 a ``small" subset of elements that are candidates to be the defective elements. In the second stage the elements selected by the first stage are tested  individually so as to determine those that are really defective. In many contexts, such as molecular biology experiments involving  the screening of library of clones \cite{knill}, two-stage algorithms are considered as practical as   non-adaptive algorithms. Indeed, in those applications,  an element must undergo an individual test in order to be confirmed as defective, even though the responses to
 previous group tests indicate with no doubt that it is defective. Therefore, the tests carried out in the second stage are not considered 
 an additional cost since the confirmatory tests should be performed anyway. 
 
 The problem of designing efficient group testing strategies consisting in a constant number of non-adaptive stages has been extended to different settings and variants of group testing and  has received much attention in the recent literature \cite{disjointGT}, \cite{sofsem}. In the present paper we prove that a phenomenon similar to the one
exhibited by classical group testing 
 occurs also in our model. Indeed, we give a trivial two-stage group testing strategy that must tolerate the same number of ``yes" responses admitted by the optimal
 adaptive algorithm. This result relies on an existential result proved for  a new variant of the well known $(p,d)$-cover free families \cite{dyry}.

  In Section 2, we present the lower bound for the adaptive case and give an algorithm that asymptotically achieves this bound. In Section \ref{sec:cover}, we first recall the  definitions of $d$-separable families, $d$-cover free families and $(p,d)$-cover free families, and describe the existing relationship between these
  combinatorial structures and classical group testing. Then, in Section \ref{sec:new_var}, we introduce
 our variants of these families which represent our main combinatorial tools.
 In Section \ref{sec:non_adaptive}, we  consider the non-adaptive scenario and derive a lower bound on the number of ``yes" responses
 that must be tolerated by any non-adaptive algorithm that uses at most a certain number $t$ of tests. This lower bound is a consequence of
 an upper bound we prove in Section \ref{sec:lower_pd} on the size of our variant of $(p,d)$-cover free families.  In Section \ref{sec:upper_pd} we give an existential result for these families based on the probabilistic method. For $p=1$, this result shows that there
 exist non-adaptive algorithms achieving  bounds which are very close to the lower bound. In Section \ref{sec:explicit}, we give an explicit construction for  our variant of $d$-cover free families which achieves the same asymptotic bound of the construction of Section \ref{sec:upper_pd}. In Section \ref{sec:two_stage}, we consider trivial two-stage
 group testing and give an algorithm that admits the same asymptotic number of positive responses of the optimal adaptive algorithm of Section 2.
 This result is based on the existential result for our variant of $(p,d)$-cover free families of Section \ref{sec:upper_pd}.

\section{Adaptive group testing}\label{sec:adaptive}
In this section we deal with the case when tests are performed adaptively by looking at the feedbacks of already performed  tests.
For the purpose of our analysis, we need to introduce the following definition.
\begin{definition}\label{def:adaptive}
Let $t$, $n$, $d$ be positive integers with $n\geq d\geq 1$, and let $O$ be a set of $n$ elements containing at most $d$ defective elements. Moreover,
let  $\A$ be a group testing strategy that finds all defective items in $O$ by at most $t$ tests. We denote by $y_\A(d,n,t)$ the maximum number of positive responses that occurs during the search process performed by $\A$, where the maximum is taken over all possible  subsets of up to $d$ defectives. The minimum value of $y_\A(d,n,t)$  is denoted by $y(d,n,t)$, where the minimum is taken over all group testing algorithms that use at most $t$ tests to find all defectives in $O$. 
\end{definition}
Notice that $y(d,n,t)$ represents the minimum number of positive responses that must be admitted in order to find up to $d$ defectives in a set of $n$ elements by at most $t$ tests.  
The following lemma is quite straightforward.
\begin{lemma}\label{lemma1}
Let $t$, $n$, $d$ be positive integers with $n\geq d\geq 1$. Then,
$y(d,n,t) \geq d.$
\end{lemma}
\proof
Suppose by contradiction that $y(d,n,t)< d$. Then, in the case when the number of defectives is exactly $d$,
there would be at least one defective element which either is never tested or appears
only in groups that contain also other defective elements. In both cases, the algorithm could not decide whether this element is  defective or not.  
This is due to the fact that the algorithm does not know a priori whether the number of defectives is $d$ or it is smaller than $d$.
\qed

In order to derive a lower bound on $y(d,n,t)$, we describe the search process by a binary tree where each internal node corresponds to a test and each leaf to one of the possible outcomes of the algorithm. For each internal node, its left branch is labelled with 0 and corresponds to  a negative response, while its right branch is labelled with 1 and corresponds to a positive response.  A path from the root to a leaf $x$ represents the sequence of tests performed by the algorithm when the set of defective items is the one associated with $x$. Obviously, for an input set of size $n$ that contains $d$ defective elements, a group testing strategy is successful if and only if the corresponding tree has  ${n \choose d}$  leaves. Let us denote by $y$ the maximum number of ``yes" responses in the whole sequence of test responses. Each root-to-leaf path can be represented by the binary vector  whose entries are the labels of the branches along the path taken in the order they are encountered starting from the root.  Since each path that starts from the root and ends in a leaf must contain at most $y$ branches labelled with 1,  the number of such binary vectors is smaller than or equal to  $\sum_{i=0}^y{t\choose i}$. 
Since the number of leaves cannot be larger than the upper bound on the number of root-to-leaf paths,  it holds

\begin{equation}\label{eq:lowerAdap1}
\sum_{i=0}^y{t\choose i}\geq {n\choose d}.
\end{equation}
The above bound obviously  holds also in the case when $d$ is an upper bound on the number of defective elements.
% since in that case the number of possible outcomes is $\sum_{k=1}^d{n\choose k}\geq {n\choose d}$.

Inequality  (\ref{eq:lowerAdap1}) allows to derive a lower bound on $y(d,n,t)$. In order to obtain the desired bound, we make use of the following 
lemma which establishes an upper bound on the binary entropy $H(\frac ab)=-\frac ab \log \frac ab-(1-\frac ab)\log(1-\frac ab)$, for any $a$ and $b$ such that $0 < a< b$.
In the following, unless specified differently, all logarithms are in base 2.

\begin{lemma}\label{lemma:ent}
Let $a$ and $b$ such that $0 <a<b$. It holds $$H\left(\frac ab\right)\leq \frac ab \log\left( \frac {eb}a\right).$$
\end{lemma}
\proof
By the definition of binary entropy, one has that
\begin{eqnarray}
H\left(\frac ab\right)&=&\frac ab \log \frac ba+\left({b-a\over b}\right)\log\left({b\over b-a}\right)\cr
&=&\frac ab \log \frac ba+{1\over b}\cdot\log\left(1+{a\over b-a}\right)^{b-a}\cr
&\leq&\frac ab \log \frac ba+{1\over b}\cdot \log   e^a,
\end{eqnarray}
from which the upper bound in the statement of the lemma follows.
\qed

Below we will often resort to the following
well known inequalities on the binomial coefficient
 \begin{equation}\label{eq:binomial_low}
{N \choose m}\geq  \left( \frac{N}{m}\right)^{m},
 \end{equation}
\begin{equation}\label{eq:binomial_upp}
 {N \choose m}\leq \left( \frac{e N}{m}\right)^{m},
 \end{equation}
where $e$ denotes the Neper's constant $e = 2,71828\ldots $.

\begin{theorem}\label{thm:lowerAdapt}
Let $t$, $n$, $d$ be positive integers with $n\geq d\geq 1$. It holds that
$$y(d,n,t) > \max\left\{d\,, \, {d\log \left( \frac{n}{d}\right)\over\log \alpha}\right\},$$
where $\alpha= 4$ if $y(d,n,t)>t/2$, and $\alpha=\frac {et}{y(d,n,t)}\leq {et\log \left(\frac {et}d\right)\over{d\log \left( \frac{n}{d}\right)}}$
 if $y(d,n,t)\leq t/2$.
\end{theorem}
\proof
Let $y$ denote the maximum number of positive responses admitted by an adaptive group testing algorithm
that uses at most $t$ tests to find up to $d$ defectives.
By inequality (\ref{eq:lowerAdap1}) we have that $\sum_{i=0}^y{t\choose i}\geq {n\choose d}$.

First we consider the case $y\leq t/2$. Stirling approximation implies the following well known inequality \cite{fg}
\begin{equation}\label{eq:stir}\sum_{i=0}^{\ell}{m\choose i}\leq 2^{mH(\ell/m)}.\end{equation}
 where $\ell/m  \leq 1/2$. By setting $m=t$ and $\ell=y$ in inequality (\ref{eq:stir}), we get 
\begin{equation}\label{eq:leftLowAdap1}
\sum_{i=0}^y{t\choose i}\leq 2^{tH(y/t)}.
\end{equation}

Lemma \ref{lemma:ent} implies that $H(\frac yt)\leq \frac yt \log \frac {et}y$,
from which one has that
\begin{equation}\label{eq:leftLowAdap2}
\sum_{i=0}^y{t\choose i}\leq 2^{y\log \frac {et}y}.
\end{equation}

The lower bound on the binomial coefficients in (\ref{eq:binomial_low}) implies that
\begin{equation}\label{eq:rightLowAdap1}
{n\choose d}\geq \left( \frac{n}{d}\right)^{d}.
\end{equation}

Therefore, inequalities (\ref{eq:lowerAdap1}), (\ref{eq:leftLowAdap2}), and (\ref{eq:rightLowAdap1}) imply that,  for $y\leq t/2$,
\begin{equation}\label{eq:lowAdap2}
2^{y\log \left(\frac {et}y\right)}\geq \left( \frac{n}{d}\right)^{d},
\end{equation}
from which one has that

\begin{equation}\label{eq:lowAdap4}
y\geq {d\log \left( \frac{n}{d}\right)\over\log \left(\frac {et}y\right)}. 
\end{equation}

Now let us turn our attention to the case when $y>t/2$. In this case the bound follows from the information theoretic lower bound.
One has that

\begin{equation}\label{eq:lowAdap5}
y>t/2\geq \frac12\left\lceil\log {n\choose d}\right\rceil.
  \end{equation}

Inequalities (\ref{eq:rightLowAdap1}) and  (\ref{eq:lowAdap5})  imply that

\begin{equation}\label{eq:y>}
y\geq  \frac {d}{2 }\log  \left( \frac{n}{d}\right).
\end{equation}
The lower bound in the statement of the theorem is obtained by taking the maximum between the lower bound in Lemma \ref{lemma1} and either  lower bound  (\ref{eq:lowAdap4}) or lower bound (\ref{eq:y>}), according to whether  $y\leq t/2$ or   $y>t/2$. The term $\alpha$ in the bound of the theorem is equal to 4 when $y>t/2$, and is equal to $\frac {et}y$ when $y\leq t/2$. In this latter case we limit from above $\alpha$  by  applying lower bound (\ref{eq:lowAdap4}) to $y$ in  the expression of $\alpha$, thus getting
$\alpha=\frac {et}y\leq{et\log \left(\frac {et}y\right)\over{d\log \left( \frac{n}{d}\right)}}$, which by the lower bound in Lemma \ref{lemma1} is at most ${et\log \left(\frac {et}d\right)\over{d\log \left( \frac{n}{d}\right)}}$.
\qed

\subsection{An asymptotically optimal algorithm}
Now we present an algorithm that almost attains the lower bound of Theorem \ref{thm:lowerAdapt}.

The algorithm is designed after  Li's stage group testing algorithm \cite{li}. While Li's analysis aims at minimizing the total number of tests, 
our algorithm performs a number of tests that depends on the  number of positive responses admitted by the algorithm. 

The algorithm works as follows.
The tests are organized in stages in such a way that each stage tests a collection of disjoint subsets that form a partition of the search space. At stage $i$ the search space is partitioned into $g_i\geq d$ groups, $g_i-1$ of which have size $k_i$, while the remaining one might have size smaller than $k_i$. The elements in the subsets that test negative are discarded, while those in the subsets that test positive are grouped together to form the new search space.
Notice that the tests in each stage can be performed in parallel. Let $f$ denote the total number of stages. Notice that in stage $i$, $i=1,\ldots, f$,   the defective elements are contained in at most $d$ of the $g_i$ groups and therefore, after this stage, the search space consists of  at most $dk_i$ elements. The algorithm is successful if and only if after stage $f$ the search space contains only the defective elements. This is insured by setting $k_f=1$.

Let us ignore for the moment the integral constraints. 
 The total number of tests performed by the algorithm is  

\begin{equation}\label{eq:upperAdapt1}
t=\sum_{i=1}^f g_i \leq  \frac n {k_1}+ \frac {dk_1}{k_2} +  \frac {dk_2}{k_3}+\ldots+ \frac {dk_{f-2}}{k_{f-1}}+dk_{f -1}.\end{equation}
As observed before, in each stage at most $d$ groups test positive and consequently, the total number of positive responses is upper bounded by $fd$. Obviously,  the minimum is attained for $f=1$, i.e, in the case when the algorithm consists in a single stage that tests each element individually.  Therefore, it trivially holds 
\begin{equation}\label{eq:dnn}
y(d,n,n)= d.
\end{equation} 

If we fix the number of stages $f$, the values of the $k_i$'s do not affect the upper bound on the number of positive responses (as far as $g_i= \frac {dk_{i-1}}{k_i}\geq d$, i.e., $k_{i-1}\geq k_i$).  Therefore, we choose the values of $k_1,\ldots,k_{f-1}$ which minimize  the upper bound on  $t$. As shown in \cite{li}, the minimum value of the right-hand side of (\ref{eq:upperAdapt1}) is attained for $k^*_i=\left(\frac {n} {d}\right)^{\frac {f-i}{f}}$, $i=1,\ldots,f-1$. 
%We set $k^*_i$ to be
%the number of elements in the groups tested at stage $i$, for $i=1,\ldots,f$.
As a consequence, we have $g_1=\left \lceil  \frac {n}{k^*_{1}}\right\rceil$ and $g_i=d\left \lceil  \frac {k^*_i}{k^*_{i+1}}\right\rceil$, for $i=2,\ldots, f$. In each stage, the
 number  of tests  is  at most $d\!\left \lceil  (\frac nd)^{\frac 1f}\right\rceil$, and consequently, the total number of tests is
$$
t\leq f   d \left(\frac {n} {d}\right)^{\frac {1}{f}}+fd-1.
$$
The above upper bound on $t$ implies 

\begin{equation}\label{eq:upperAdapt2}
f\leq\frac {\log(\frac{n}{d})} {\log(\frac{t}{fd}-1+\frac{1}{fd})}=\frac {\log(\frac{n}{d})} {\log(\frac{t+1}{fd}-1)}.
\end{equation}
Since the maximum number of positive responses is $fd$, we set $y_\A(d,n,t) = fd$ and have that inequality (\ref{eq:upperAdapt2}) implies that

\begin{equation}\label{eq:dnt}
y_\A(d,n,t) \leq 
\frac {d\log(\frac{n}{d})} {\log(\frac{t+1}{y_\A(d,n,t)}-1)}.
%<\frac {d\log(\frac{n}{d})} {\log(\frac{t}{y_\A(d,n,t)}-1)}.
\end{equation}

If the number   $y_\A(d,n,t)$ of ``yes" responses tolerated by the algorithm is larger than $\frac t  3$ and $t<n$, then, 
in place of the above described algorithm,   we use  Hwang's algorithm \cite{hwang} for 
classical group testing. This algorithm   performs at most $d-1$ more tests  than the information theoretic lower bound and
therefore we have 
\begin{equation}\label{eq:hwang}
y_\A(d,n,t) \leq t\leq\left\lceil\log {n\choose d}\right\rceil+d-1.
\end{equation}

The bounds in the statement of the following theorem  
follow from (\ref{eq:hwang}), (\ref{eq:dnt}), and (\ref{eq:dnn}).
The lower bound on $\gamma=\frac{t+1}{y_\A(d,n,t)}-1$ in the statement of the theorem is obtained by observing that, by upper bound (\ref{eq:dnt}), it holds
$$\gamma=\frac{t+1}{y_\A(d,n,t)}-1\geq \frac{(t+1){\log\left(\frac{t+1}{y_\A(d,n,t)}-1\right)}}{d\log(\frac{n}{d})} -1>\frac{(t+1)}{d\log(\frac{n}{d})} -1,$$
where the last inequality is a consequence of $y_\A(d,n,t)$ being at most $\frac t 3$, from which it follows  that ${\log\left(\frac{t+1}{y_\A(d,n,t)}-1\right)}> 1$.

\begin{theorem}\label{thm:upperAdapt}
Let $t$, $n$, $d$ be positive integers with $n\geq d\geq 1$. There exists a group testing strategy $\A$ for which it holds that 
$$y_\A(d,n,t) \leq \cases{
d& if $t=n$,\cr\cr
 \left\lceil\log {n\choose d}\right\rceil+d& if $t<n$ and $y_\A(d,n,t)>t/3$,\cr\cr
\frac {d\log(\frac{n}{d})} {\log\gamma} &if $t<n$ and $y_\A(d,n,t)\leq t/3$,\\}$$
where $\gamma=\frac{t+1}{y_\A(d,n,t)}-1>\frac{(t+1)}{d\log(\frac{n}{d})} -1
%{t+1 \over d\log\left(\frac{n}{d}\right)}-1
$.
\end{theorem}
If we consider the case when  more than $1/3$ of the tests may receive a ``yes" response, then it is  immediate to see that the algorithm of Theorem \ref{thm:upperAdapt} asymptotically attains the lower bound of Theorem \ref{thm:lowerAdapt}.

Let us consider the case when at most $1/3$  of the total number of tests are allowed to receive a ``yes" response. Notice that the upper bounds of Theorem \ref{thm:upperAdapt} translate into  upper bounds on the number of tests that suffice to find up to $d$ defective elements by a group testing  algorithm that admits at most $y=y_\A(d,n,t)$ ``yes" responses. Seen in this way, Theorem \ref{thm:upperAdapt} implies that there exists  an algorithm  that uses 
\begin{equation}\label{eq:16_6_15_1}
t\leq y2^{d\log\left(\frac nd\right)\over y}+y-1\end{equation}
 tests, where $y\leq\frac t3$ is the maximum number of positive responses admitted by the algorithm.
Similarly, the lower bounds stated by  Theorem \ref{thm:lowerAdapt} translate into lower bounds on the number of tests performed by any group testing algorithm that admits at most a certain number $y$ of positive responses. If we consider algorithms that allow at most $1/3$ of the tests to yield a ``yes" response, Theorem \ref{thm:lowerAdapt}  implies that any such algorithm performs at least  
\begin{equation}\label{eq:16_6_15_2}
t\geq \frac 1 e y2^{{d\log \left(\frac nd\right)   \over y}}
\end{equation} tests. 
The ratio between the upper bound (\ref{eq:16_6_15_1})  and the  lower bound  (\ref{eq:16_6_15_2}) is a constant, and as a consequence, the algorithm of Theorem \ref{thm:upperAdapt} is asymptotically optimal.

\section{Cover-free families and group testing  }\label{sec:cover}
 In this section, we describe the existing relationship  between non-adaptive group testing and well known combinatorial structures such as $d$-separable families,
 $d$-cover free families  and $(p,d)$-cover free families.
 We recall that a group testing algorithm is said to be {\em non-adaptive} if all tests must be decided beforehand without looking at the responses of previous tests.

In the following, for any positive integer $m$, we denote by $[m]$ the set of integers $\{1,\ldots,m\}$ and by $[m]_k$, $1\leq k\leq m$, the set of all $k$-element subsets of $[m]$.

There exists a correspondence between non-adaptive group testing algorithms for  input sets of size $n$ and  families of $n$ subsets. Indeed, given a family  $\F=\{F_1,\ldots,  F_n\}$ with $F_i\subseteq [t]$, we design a non-adaptive group testing strategy as follows.  We denote the elements in the input set by the integers in $[n]=\{1,\ldots, n\}$ and for $i=1,\ldots, t$, define the group $T_i=\{j\,:\, i\in F_j\}$. Obviously,     $T_1,\ldots,T_t$ can be tested in parallel and therefore  the resulting algorithm is non-adaptive. Conversely, given a non-adaptive group testing strategy for an input set of size $n$ that tests $T_1,\ldots,T_t$, we define a family $\F=\{F_1,\ldots,F_n\}$  by setting $F_j=\{i\in [t]\,:\, j\in T_i\}$, for $j=1,\ldots, n$. 
Equivalently, any non-adaptive group testing algorithm for an input set of size $n$ that performs $t$ tests corresponds to a binary code of length $t$ and size $n$. This is due to the fact that any family of size $n$ on the ground set $[t]$ can be represented by the binary code of length $t$ whose codewords are the characteristic vectors of the members of the family. Given such a binary code  $\C=\{\cc_1,\ldots,\cc_n\}$, one has that  $j$ belongs to  pool $T_i$ if and only if the $i$-th entry $\cc_j(i)$ of $\cc_j$ is equal to 1.
  
A non-adaptive group testing strategy is successful if and only if the corresponding family is a $\dd$-{\em separable} family, i.e., a family in which the unions of up to $d$ members are pairwise distinct \cite{dh2,dh1}. To see this, 
let us represent the test responses by a binary vector whose $i$-th entry is equal to 1 if and only if $T_i$ tests   positive. We call this vector the {\em response vector}. Notice that the response vector  is the characteristic vector of the union of the members of the family associated with the  defective elements. In the binary code representation, this is equivalent to saying that the response vector is the $OR$ of the codewords associated with the defective elements.
Therefore, the set of the defective elements is univocally identified if and only if the union of up to $d$ members of the family are pairwise distinct, that is,  if and only if the family is $\dd$-separable. The reader is referred to \cite{dh2,dh1} for a detailed account on these issues.

In spite of the equivalence between separable families and non  adaptive group testing strategies, typically in the literature the design of non-adaptive algorithms is based on families satisfying a  slightly stronger property that allows for a more efficient decoding algorithm 
 to obtain the set of defectives from the test responses. 
These families satisfy the 
 property  that no member of the family is contained in the union of any other $d$ members. Families with this property are called $d$-{\em cover free} families \cite{erff}, whereas  the corresponding binary codes are said to be $d$-{\em superimposed} or $d$-{\em disjunct} 
 \cite{dh2}, \cite{dh1}, \cite{dyry}, \cite{ks}.  Such  codes have the property that for each codeword $\cc$ and any other $d$ codewords $\cc_{j_1},\ldots, \cc_{j_d}$ there exists an index $i$ such that $\cc$ has the $i$-th entry equal to 1, whereas all of  $\cc_{j_1},\ldots, \cc_{j_d}$ have the $i$-th entry equal to 0. Given two binary vectors $\cc_1$ and $\cc_2$ of length $t$, we say that $\cc_2$ 
 {\em covers} $\cc_1$ if for any $i\in[t]$, $\cc_1(i)=1$ implies that $\cc_2(i)=1$.
 By using this terminology, we say that a code is $d$-superimposed (or $d$-disjunct) if and only if no codeword is covered by the Boolean $OR$ of any other $d$ columns. A consequence of this property is that any codeword associated with a regular (e.g., non defective) element is not
 covered by the  response vector. Therefore, it is possible to recover the set of the defective elements by simply comparing the response vector  with each codeword. On the other hand,
if we use an algorithm based on a $\dd$-separable family then, in order to  obtain the set of the defective elements,  we need to examine all subsets of up to $d$ codewords.

The $d$-cover free families are a particular case of  the $(p,d)$-{\em cover free families} introduced by D'yachkov and Rykov in \cite{dyry} under
the name of superimposed $(d,n,p)$-codes, where $n$ denotes the size of the family.
A $(p,d)$-cover free family is a family 
such that the union of any $p$ members of the family is not contained in the union of any other $d$ members of the family.
For $p=1$, $(p,d)$-cover free families are equivalent to $d$-cover free families. 
Analogously to what happens with $d$-cover free families,  $(p,d)$-cover free families can be associated with  non-adaptive group testing algorithms. However, these algorithms do not guarantee to determine  exactly  all defectives but allow only to
obtain a subset of at most $p+d-1$ elements containing all defective elements. Indeed, given a response vector ${\bf z}$, there might be up to $p+d-1$ members of the families whose characteristic vectors are covered by ${\bf z}$. This is due to
the fact that for any possible subset of up to $d$ defective elements there are at most $p-1$ other elements such that
the members of the families corresponding to these elements are contained in the union of the members associated with the defective elements.
The authors of \cite{dgv} used a $(d,d)$-cover free family to design the first stage of their two-stage algorithm. This stage 
allows to determine a subset of up to $2d-1$ elements including all defective elements. The elements in this subset are individually
tested during the second stage in order to find out which ones of them are defective.

As a matter of fact, the authors of \cite{dgv} based their algorithms on $(k,m,n)$-selectors, a combinatorial structure satisfying 
a slighter stronger property than that of $(p,d)$-cover free families. Their existential
 result for this combinatorial structure implies
 that there exists a $(p,d)$-cover free family of
size $n$ on a ground set of size 
\begin{equation}\label{eq:siam}
t<{e(p+d)^2\over p}\ln \frac n{p+d}+ {e(p+d)(2(p+d)-1)\over p}.
\end{equation}

\subsection{New variants of separable and cover-free families} \label{sec:new_var}
In this section we  introduce 
  variants of separable and cover-free families that can be used to derive upper and lower bounds for the group testing problem we are considering.
  
Let $\F=\{F_1,\ldots,F_n\}$ be a family of subsets of $[t]=\{1,\ldots,t\}$. We will refer  to the set $[t]$ as the {\em ground set} of the family. For a positive integer $k\leq t$, a family  $\F=\{F_1,\ldots,F_n\}$ of subsets of $[t]$ is said to be $k$-{\em uniform}, if $|F_i |=k$, for $i=1,\ldots,n$.

Given a family $\F=\{F_1,\ldots,F_n\}$, the corresponding group testing algorithm must admit a number of positive responses which is as large as the size of the largest union of up to $d$ members of the family. Indeed, let ${j_1},\ldots, {j_m}$, with $m\leq d$, be the defective elements. A group $T_i$ intersects 
 $\{j_1,\ldots, j_m\}$ if and only if $i\in  F_{j_1}\cup \ldots\cup F_{j_m}$. Therefore, the number of positive responses is equal to $|F_{j_1}\cup \ldots\cup F_{j_m}|$. 
 By the above argument, a  non-adaptive group testing strategy that uses $t$ tests and admits at most $s$ positive responses is equivalent to the following notion 
 of $\cup_{\leq s}$ $\dd$-separable family.
 
\begin{definition}\label{defi:cfLimU}
Let $d$, $s$, and $t$, $s\leq t$, be positive integers.
We say that a family $\F$ on the ground set $[t]$ 
is a $\cup_{\leq s}$ $\dd$-{\em separable} family if the unions  of up to $d$ members of $\F$ are all distinct, and the union of any $d$ members of $\F$ has size at most $s$.
The maximum cardinality  of a  $\cup_{\leq s}$ $\dd$-separable family on the  ground set $[t]$ is 
denoted by $n_{sep}(d,\cup_{\leq s},t)$.
\end{definition}
Analogously to what happens in classical group testing, cover free families allow to decode the response vector much more efficiently.
Therefore, we introduce the following definition.
 \begin{definition}\label{defi:cfLimU}
Let $d$, $s$, and $t$, $s\leq t$, be positive integers.
We say that a family $\F$ on the ground set $[t]$
is a $\cup_{\leq s}$ $d$-{\em cover free} family if no member of $\F$ is contained in the union of other $d$ members of $\F$, and the union of any $d$ members of $\F$ has size at most $s$.
The maximum cardinality  of a $\cup_{\leq s}$ $d$-cover free family on the  ground set $[t]$ is 
denoted by $n_{cf}(d,\cup_{\leq s},t)$.
\end{definition}
It is immediate to see that  $\cup_{\leq s}$ $d$-cover free families are  $\cup_{\leq s}$ $\dd$-separable families, and consequently, 
existential results for the former families apply also to the latter families. 
The following theorem shows that upper bounds on the maximum cardinality of $\cup_{\leq s}$ $(d-1)$-cover free families can be used to
derive  upper  bounds on the maximum size of $\cup_{\leq s}$ $\dd$-separable families.

 \begin{theorem}\label{thm:almostequivalent}
Let $d$, $s$, and $t$, $s\leq t$, be positive integers.
Any $\cup_{\leq s}$ $\dd$-separable family is $\cup_{\leq s}$ $(d-1)$-cover free.
\end{theorem}
\proof
First we show  that any $\dd$-separable family is a $(d-1)$-cover free family. This relation was noted by Kautz and Slingleton \cite{ks} and is quite simple to see. Indeed, suppose by contradiction that a $\dd$-separable family is not $(d-1)$-cover free. As a
consequence, there exist $d$ members of the family $F_1,F_2,\ldots,F_{d}$ such that $F_{d}\subseteq F_1\cup \ldots \cup F_{d-1}$,
 and therefore, it holds $\bigcup_{i=1}^d F_i=\bigcup_{i=1}^{d-1}F_i$ thus contradicting the fact that the family is $\dd$-separable.
Moreover, for any $d$ members $F_1,F_2,\ldots,F_{d}$, it holds  $|\bigcup_{i=1}^{d-1}F_i|<|\bigcup_{i=1}^d F_i|\leq s$, thus proving that the family is $\cup_{\leq s}$ $(d-1)$-cover free.
\qed

If we are not interested  in determining exactly which elements  are defective  but only in confining the defective elements inside a reasonably 
small subset, then the following definition provides an useful combinatorial tool.

\begin{definition}\label{defi:pd_cover_free}
 Let  $p, d$, $s$, and $t$, $s\leq t$, be positive integers.
 We say that a family $\F$ on the ground set $[t]$
 is a $\cup_{\leq s}$ $(p,d)$-{\em cover free} family  if the union of any $p$ members of $\F$ is not contained in the union of other $d$ members of  $\F$,   and the union of any $d$ members of $\F$
 has size at most $s$,
 The maximum cardinality  of  a $\cup_{\leq s}$ $(p,d)$-cover free family on the ground set $[t]$ will be denoted   by $n_{cf}(p,d,\cup_{\leq s},t)$
\end{definition}

The non-adaptive algorithm designed after a $\cup_{\leq s}$ $(p,d)$-cover free family has the pro\-perty that at most $s$ pools test positive and that at most $p-1$ non defective elements cannot be classified as such. Indeed, there are at most $p-1$ non defective elements 
 that appear only in pools containing  one or more defective elements. In other words, the response vector has weight at most $s$ and covers
 at most $p+d-1$ codewords of the binary code associated with the family, that is,  at most $p-1$ codewords in addition to those associated with the defective elements.
 
In Section \ref{sec:two_stage},   a $\cup_{\leq s}$ $(p,d)$-cover free family is employed to design the pools tested in the first stage of our trivial two-stage algorithm so that at most $d+p-1$ elements are candidates to be the defective elements after the first stage and should be
individually probed during the second stage.

\section{Non-adaptive group testing} \label{sec:non_adaptive}
In this section we present almost matching  upper and lower bounds on the number of positive responses that should be admitted by
a non-adaptive algorithm that uses at most $t$ tests to find up to $d$ defective elements in a given set of $n$ elements.
These bounds are obtained by establishing upper and lower bounds on the maximum size of $\cup_{\leq s}$ $d$-cover free families
on the ground set $[t]$.  Indeed, these bounds translate, respectively, into lower and upper bounds on the number $s$ of positive responses that might be given
to the tests. Our upper bound as well as one of our two constructions are given for the  more general
case of $\cup_{\leq s}$ $(p,d)$-cover free families. This existential result  is proved by the probabilistic method and for $p=1$ it achieves the same asymptotic bound of the construction for $\cup_{\leq s}$ $d$-cover free families given in \cite{cocoa2014}, while improving on the estimate of the  constant hidden in the asymptotic notation. The construction for $\cup_{\leq s}$ $(p,d)$-cover free families will be also employed to design the pools tested in the first stage of the two-stage algorithm of  Section \ref{sec:two_stage}. 
Our second existential result is proved directly for $\cup_{\leq s}$ $d$-cover free families. This construction exhibits the interesting feature of being an explicit construction while attaining the same  bound as the probabilistic construction.

In the following, given a non-adaptive algorithm $\A$ that  finds up to $d$ defective elements in an input set of size $n$ by at most $t$ tests, we denote by $\y_\A(n,d,t)$ the maximum  number of positive responses that may occur during the search process performed by $\A$,
 where the maximum is taken over all possible  subsets of up to $d$ defectives. Moreover, we denote by $\y(n,d,t)$ the minimum value of $\y_\A(n,d,t)$  over all non-adaptive strategies $\A$ that  find up to $d$ defective elements in an input set of size $n$ by at most $t$ tests.

\subsection{Negative Result}\label{sec:lower_pd}
%In the following, $e $ will always denote the Neper's constant $e = 2,71828\ldots $.
 \begin{theorem}\label{thm:low_p_d_cf_low}
Let $d$ and $p$ be positive integers and let $s$ and $t$ be integers such that $ s\leq t$. The maximum size of  a $\cup_{\leq s}$ $(p,d)$-cover free family on the ground set $[t]$ is $$n_{cf}(p,d,\cup_{\leq s},t)\leq 
\cases{
{t\choose \lceil t/2\rceil } &if $d=1$, $p=1$, and $t<2s$,\cr
{t\choose s} &if $d=1$, $p=1$, and $t\geq 2s$,\cr
(p+d-1)2^{\frac td} &if  $d=1<p$ or $2\leq d< 2p$, and  $t< 2s$,\cr\cr
(p+d-1)\left(\frac {et}s\right)^{\frac sd} &if $d=1<p$ or $2\leq d< 2p$, and  $t\geq 2s$,\cr\cr
p\left({et d(d+2)\over 4ps}
\right)^{\left\lceil{s\over p\lfloor d/(2p)\rfloor^2+\lfloor d/(2p)\rfloor}\right\rceil}+\frac d2+2p-2   &  if
$d\geq 2p.$\\
}
$$

\end{theorem} 
\proof
%The bound for the case $t<2s$ follows from the bound for classical $(p,d)$-cover free families in Theorem 2  of \cite{dgv}.
%The rest of the proof is devoted to the case $t\geq 2s$.
The first bound for the case $d=1$ and $p=1$ follows from the upper bound $\F\leq {t\choose \lceil t/2\rceil}$ on the size of a Sperner family $\F$ on the ground set $[t]$ with members of unlimited size, while the second bound for the case $d=1$ and $p=1$ follows  from the upper bound $\F\leq {t\choose s}$ on the size of a Sperner family $\F$ on the ground set $[t]$ and with members of size at most $s\leq t/2$.

Let us prove the bound for $2\leq d< 2p$. In this case the bound is a consequence of  Proposition 2 in \cite{dyry}. The authors of
 \cite{dyry} noticed that for any subfamily $Q$, with $|Q|\leq d$,
of a $(p,d)$-cover free family, there are at most ${d+p-1\choose d}$  subfamilies of $d$ members of the family such that the union of the $d$ members in each of these subfamilies is equal to the union of the $d$ members of $Q$. This implies that for a $(p,d)$-cover free family of size $n$, there are at least 
${{n\choose d}\over {d+p-1\choose d}}$ distinct sets that can be obtained from the union of $d$ members of the family. 
Since our $(p,d)$-cover free families  have the additional property that the union of any $d$ members of the family has size at most $s$,
the following condition must be satisfied.
\begin{equation}\label{eq:dyry1}
\sum_{i=0}^s {t\choose i}\geq{{n\choose d}\over {d+p-1\choose d}},
\end{equation}
where the sum in the left-hand side represents the maximum number of subsets of $[t]$ of size less than or equal to $s$.

For $t\geq 2s$, we bound $\sum_{i=0}^s {t\choose i}$ by exploiting  inequality (\ref{eq:leftLowAdap2}) in Section \ref{sec:adaptive}, whereas for $t< 2s$, we bound from above $\sum_{i=0}^s {t\choose i}$ by $\sum_{i=0}^t {t\choose i}$, and therefore, we have that 
\begin{equation}\label{eq:ent}
\sum_{i=0}^s {t\choose i}\leq 
\cases{
2^{s\log \frac {et}s} &if $t\geq 2s$,\cr\cr
 2^t &  if
$t< 2s.$\\
}
\end{equation}

By inequality (\ref{eq:ent}) and inequality (\ref{eq:dyry1}), one has that for $t\geq 2s$, 
\begin{equation}\label{eq:dyry2}
2^{s\log \frac {et}s}\geq{{n\choose d}\over {d+p-1\choose d}},%\geq {\left(\frac nd\right)^d\over \left(\frac {e(d+p-1)}d\right)^d},
\end{equation}
%where the last inequality follows from the upper and lower bounds  on the binomial coefficients in (\ref{eq:binomial}),
whereas for $t<2s$, it holds that
\begin{equation}\label{eq:dyry3}
2^t\geq {{n\choose d}\over {d+p-1\choose d}}.
\end{equation}
The bound (\ref{eq:dyry3}) is the same bound obtained by \cite{dgv} for the case $d<2p$.

The right-hand side of (\ref{eq:dyry1}) is equal to
$${n!\over (n-d)!d!}\cdot {d!(p-1)!\over (d+p-1)!}={n(n-1)\cdots(n-d+1)\over (d+p-1)(d+p-2)\cdots p}\geq \left({n\over d+p-1}\right)^d.$$
%The upper and lower bounds  on the binomial coefficient in (\ref{eq:binomial_low}) and (\ref{eq:binomial_upp})  imply $%{{n\choose d}\over {d+p-1\choose d}}\geq {\left(\frac nd\right)^d\over \left(\frac {e(d+p-1)}d\right)^d}$. 
 Therefore, we can lower bound the right-hand sides of (\ref{eq:dyry2}) and (\ref{eq:dyry3}) by $ \left({n\over d+p-1}\right)^d$, thus getting
 \begin{equation}\label{eq:dyry2_Bis}
2^{s\log \frac {et}s}\geq\left({n\over d+p-1}\right)^d, \,\,\,\,\mbox{for $t\geq 2s$}, 
\end{equation}
whereas for $t<2s$, it holds that
\begin{equation}\label{eq:dyry3_Bis}
2^t\geq \left(  {n\over d+p-1}\right)^d,\,\,\,\,\mbox{for $t<2s$}.
\end{equation}
The bounds for $2\leq d<2p$ in the statement of the theorem follow immediately from (\ref{eq:dyry2_Bis}) and (\ref{eq:dyry3_Bis}). 

\medskip 
Now let  us turn our attention to the case $d\geq 2p$. We assume for the moment that $d$ be a multiple of $2p$ and drop this assumption later on.
Let $\F$ be  a $\cup_{\leq s}$ $(p,d)$-cover free family on the ground set $[t]$ and let  us define the sets $G_1,\ldots,G_{d/2}$ as follows. We set $G_1$ to be the largest member of $\F$ and, for each $i=2,\ldots,d/2$, $G_i$ to be the largest set in $\{F\setminus \bigcup_{j=1}^{i-1}G_j: F\in \F\setminus\{G_1,\ldots,G_{i-1}\}\}$.  In other words, after choosing $G_1$ as the largest member of the family, we remove the elements of $G_1$ from all members of $\F\setminus \{G_1\}$ and set $G_2$ to be the largest of the resulting sets. Then, we remove the elements of $G_2$ from all unselected sets and set $G_3$ to be the largest of the sets of the form $F\setminus (G_1\cup G_2)$, for $F\in \F\setminus \{G_1,G_2\}$, and so on until $d/2$ sets are selected.
Let $\F' $ be the family obtained by removing the elements of $G_1,\ldots,G_{d/2}$ from all members of $\F\setminus \{G_1,\ldots,G_{d/2}\}$, i.e., $\F'=\{F\setminus \bigcup_{j=1}^{d/2}G_j: F\in \F\setminus\{G_1,\ldots,G_{d/2}\}\}$.
We show that the the union of any $p$ members of $\F'$  is not contained in the union of any other   $d/2$ members of the family. Suppose by contradiction that there  are $d/2 +p$ sets $F'_1,F'_2,\ldots, F'_{d/2+p}\in \F'$ such that 
 $F'_1\cup\ldots\cup F'_{p}\subseteq F'_{p+1}\cup\ldots\cup F'_{d/2+p}$. Since for $i=1,\ldots d/2+p$, it is  $F'_i= F_i\setminus \bigcup_{j=1}^{d/2}G_j$ for some set $F_i\in \F\setminus\{G_1,\ldots,G_{d/2}\}$,  
 it holds $F_1\cup\ldots\cup F_{p}\subseteq F_{p+1}\cup\ldots\cup F_{d/2+p}\cup G_1\cup\ldots\cup G_{d/2}$, thus contradicting the fact that $\F$ is $(p,d)$-cover free. Notice that it might be that the members of $\F'$ are not pairwise distinct and that some members of $\F'$ are empty. By the same argument as above one can prove that there exist at most $p-1$ sets $B_i\in \F'\cup\{G_1,\ldots,G_{d/2}\}$ such that $B_i=\emptyset$ or $B_i=B_j$ for some other member of $B_j\in \F'\cup\{G_1,\ldots,G_{d/2}\}$. If we remove these up to $p-1$ sets from $\F'$, we obtain a collection whose members are non-empty and pairwise distinct.
Let us denote by $\F''$ this collection. %the family obtained by removing from $\F'\cup\{G_1,\ldots,G_{d/2}\}$ these up to $p-1$ sets.
By construction, $\F''$ is a  $\cup_{\leq s}$ $(p,d/2)$-cover free family of cardinality larger than or equal to $ |\F|-d/2-p+1$.
 In the following, we derive an upper bound on the cardinality of $\F''$. To this aim, we exploit the  fact that the members of $\F''$ are non-empty and pairwise distinct and that $\F''$ is   $\cup_{\leq s}$ $(p,d/2)$-cover free.
  
Notice that $G_1,\ldots,G_{d/2}$ are pairwise disjoint and that $|G_1|\geq |G_2|\geq \ldots \geq |G_{d/2}|$. Moreover, 
it holds $G_i\cap F''=\emptyset$  and $ |G_{i}|\geq |F''|$, for any $i=1,\ldots,d/2$ and $F''\in\F''$.
Therefore, for any member $F''\in \F''$, one has that
\begin{equation}\label{eq:upperOfUnion}
\big|\bigcup_{i=1}^{d/2} G_{i}\cup F''\big|= \sum_{i=1}^{d/2} |G_{i}|+|F''|\geq (d/2+1) |F''|.
\end{equation} 
Since $G_1,\ldots,G_{d/2}$ are members of $\F$ and $F''$ is subset of some member of $\F$, one has that
$\big|\bigcup_{i=1}^{d/2} G_{i}\cup F''\big|\leq s$, which, along with (\ref{eq:upperOfUnion}), implies
$|F''|\leq \lfloor\frac{2s}{d+2}\rfloor$. 
Since $F''$ is an arbitrary member of $\F''$, inequality (\ref{eq:upperOfUnion}) holds for any member $F''$ of $\F''$.

% As observed in the proof of Theorem 2 of  \cite{dgv}, 
Observe that if $d$ is a multiple of $p$ then
 for any $p$ members $F_1,\ldots, F_p$ of size at most $m$
 of a $(p,d)$-cover free family, there exists a subset $A$ of  
 at most $\lceil mp/d\rceil$  elements such that $A\subseteq F_j$ for some $F_j\in \{F_1,\ldots,F_p\}$ and $A\not\subseteq F$
 for any member $F$ of the family such that $F\not\in \{F_1,\ldots,F_p\}$.
 Indeed, if otherwise it would be possible to partition each of $F_1,\ldots,F_p$ into  $d/p$ subsets of size at most
 $\lceil mp/d\rceil$ each of which is contained in a member of the family different from $F_1,\ldots,F_p$. This would  imply that there
 exist $\leq d$ members of the family that contain all elements of $F_1\cup\ldots\cup F_p$, thus contradicting the hypothesis of
 the family being a $(p,d)$-cover free family. 
Since, by assumption, $d/2$ is a multiple of $p$, we can apply this observation to our $(p,d/2)$-cover free family $\F''$.
We proved that all members of $\F''$ have size at most $\lfloor\frac{2s}{d+2}\rfloor$, therefore the above observation  implies
that, for any $p$ members $F''_1,\ldots F''_p$ of  $\F''$,
 there exists a set $A$ of size at most $\lceil 4sp/(d(d+2))\rceil$
such that $A\subseteq F''_j$ for some $F''_j\in \{F''_1,\ldots,F''_p\}$ and $A\not\subseteq F''$
 for any member $F''$ of the family different from $F''_1,\ldots,F''_p$.
Now let us form $\lfloor |\F''|/p\rfloor$ pairwise disjoint subfamilies $\F''_1,\ldots,\F''_{\lfloor |\F''|/p\rfloor}$ of $\F''$
each consisting of $p$ members of $\F''$.
By the above argument, for each such  a subfamily $\F''_i$ there exists a subset $A_i$ of at most $\lceil 4sp/(d(d+2))\rceil$ elements such that $A_i$ is entirely contained in
some member of $\F''_i$ and is not contained in any member of $\F''_j$, for $j\neq i$. It follows that the
family $\{A_1,\ldots, A_{\lfloor |\F''|/p\rfloor}\}$ is a Sperner family, i.e., an antichain. The following celebrated inequality,  known under the name of
LYM inequality, establishes a relationship between the cardinalities of the members of a Sperner family $\G$ and the  size $m$ of the ground set  of the family. 

\begin{equation}\label{eq:LYM}
\sum_{G\in\G} \frac 1 {{m\choose |G|}}\leq 1.
\end{equation}
Since $\{A_1,\ldots, A_{\lfloor |\F''|/p\rfloor}\}$ is a Sperner family on the ground set $[t]$, LYM inequality implies
\begin{equation}\label{eq:LYM1}
\sum_{i=1}^{\lfloor |\F''|/p\rfloor}\frac 1 {{t\choose |A_i|}}\leq 1.
\end{equation}
Moreover, $A_1,\ldots, A_{\lfloor |\F''|/p\rfloor}$  have size at most 
$\lceil 4sp/(d(d+2))\rceil$ which, by the assumption $d\geq 2p\geq 2$, is at most $\lceil s/2\rceil\leq \lceil t/2\rceil$. Therefore,
one has that ${t\choose  |A_i|}\leq {t\choose \lceil 4sp/(d(d+2))\rceil}$, for $i=1,\ldots,\lfloor |\F''|/p\rfloor$. It follows that   the left-hand side of (\ref{eq:LYM1}) is larger than or equal to $\frac {\lfloor |\F''|/p\rfloor} {{t\choose \lceil 4sp/(d(d+2))\rceil}}$ thus implying
$\lfloor |\F''|/p\rfloor\leq {t\choose\lceil 4sp/(d(d+2))\rceil}$,
from which 
\begin{equation}\label{eq:F''}
|\F''|\leq p{t\choose \lceil 4sp/(d(d+2))\rceil}+p-1.
\end{equation}
Since $|\F|\leq |\F''|+d/2+p-1$,  inequality (\ref{eq:F''}) implies
\begin{equation}\label{eq:multiple}
|\F|\leq p{t\choose \lceil 4sp/(d(d+2))\rceil}+d/2+2p-2.\end{equation}
Now let us drop the assumption that $d$ is a multiple of $2p$. Observe that $d\geq 2p\lfloor d/(2p)\rfloor$ and therefore, one has that
$$n_{cf}(p,d,\cup_{\leq s},t)\leq n_{cf}(p,2p\lfloor d/(2p)\rfloor,\cup_{\leq s},t).$$
We upper bound $n_{cf}(p,2p\lfloor d/(2p)\rfloor,\cup_{\leq s},t)$ by using  (\ref{eq:multiple})  with $d$ replaced by $2p\lfloor d/(2p)\rfloor$, thus
obtaining 
$$n_{cf}(p,d,\cup_{\leq s},t)\leq p{t\choose \big \lceil {s\over p\lfloor d/(2p)\rfloor^2+\lfloor d/(2p)\rfloor}\big\rceil}+p\lfloor d/(2p)\rfloor+2p-2.$$
The bound for $d\geq 2p$ in the statement of the theorem follows from applying  the upper bound 
in  (\ref{eq:binomial_upp}) to the binomial coefficient in the above inequality.
\qed
By setting $p=1$ in the bound of Theorem \ref{thm:low_p_d_cf_low}, we obtain the following upper bound on the maximum size of  $\cup_{\leq s}$ $d$-cover free families. 
\begin{corollary}\label{cor:low_cfLimUn1}
Let $d\geq 1$, $s$ and $t$, $s\leq t$, be positive integers. The maximum size of  a $\cup_{\leq s}$ $d$-cover free family on the ground set $[t]$ is 
$$n_{cf}(d,\cup_{\leq s},t)\leq 
\cases{
{t\choose \lceil t/2\rceil} &if $d=1$ and $t<2s$,\cr
{t\choose s} &if $d=1$ and $t\geq 2s$,\cr
\left({et d(d+2) \over 4s}
\right)^{\left\lceil\frac{s}{\lfloor d/2\rfloor^2+\lfloor d/2\rfloor}\right\rceil}+\frac d2 &if $d\geq 2$.\\
}$$
\end{corollary}

The following theorem establishes an upper bound on the maximum size of $\cup_{\leq s}$ $\dd$-separable families on the ground set $[t]$.% is a consequence of Theorem  \ref{thm:almostequivalent} and Corollary \ref{cor:low_cfLimUn1}.
\begin{theorem}\label{thm:low_sepLimUn1}
Let $d\geq 1$, $s$ and $t$, $s\leq t$, be positive integers. The maximum size of a $\cup_{\leq s}$ $\dd$-separable family on the ground set $[t]$ is 
$$n_{sep}(d,\cup_{\leq s},t)\leq 
\cases{
2^{2s-1}  &if $d=1$ and $t<2s$,\cr
2^{s\log (et/s)} &if $d=1$ and $t\geq 2s$,\cr
2^{(t+1)/2}+1 &if $d=2$ and $t<2s$,\cr
2^{\frac s2\log\left(\frac{et}s\right)+\frac12}+1 &if $d=2$ and $t\geq 2s$,\cr
\left({et (d^2-1)\over 4s}
\right)^{\left\lceil\frac{s}{\lfloor {d-1}/2\rfloor^2+\lfloor {d-1}/2\rfloor}\right\rceil}+\frac {d-1}2 &if $d\geq 3$.\\
}$$
\end{theorem}
\proof
The bounds for   $d=1$  follow from the fact that the members of  a $\cup_{\leq s}$ $\dd$-separable family are pairwise distinct and have size at most $s$. As a consequence, it holds $n_{sep}(d,\cup_{\leq s},t)\leq \sum_{i=1}^{s}{t\choose i}$. For $d=1$ and  $t<2s$, we bound $\sum_{i=1}^{s}{t\choose i}$ by $2^t$ , thus obtaining $n_{sep}(d,\cup_{\leq s},t)\leq 2^t\leq2^{2s-1}$.  For   $d=1$ and  $t\geq 2s$, we bound $\sum_{i=1}^{s}{t\choose i}$ by  exploiting inequality (\ref{eq:leftLowAdap2}) which 
implies $\sum_{i=1}^{s}{t\choose i}\leq 2^{s\log (et/s)}$, and consequently,  $n_{sep}(d,\cup_{\leq s},t)\leq 2^{s\log (et/s)}$. 

The bound
for $d=2$ and $t<2s$ follows directly from Lindstorm's bound   \cite{dh2}  which limits from above  the size of $\bar 2$-separable families on the ground set $[t]$ by $1+  2^{(t+1)/2}$.
For $d=2$ and $t\geq 2s$, the stated bound  follows from observing that the unions of any two members of a 
 $\cup_{\leq s}$ $\bar 2$-separable family are distinct and have size smaller than or equal to $s$. Therefore,
 it must be ${n_{sep}(d,\cup_{\leq s},t)\choose 2}\leq \sum_{i=1}^s{t\choose i}$. Then, the stated bound for  $d=2$ and $t\geq 2s$
 follows from inequality (\ref{eq:leftLowAdap2}).
 
 The bound for $d\geq 3$ follows immediately from Theorem  \ref{thm:almostequivalent} and from  the upper bound stated by  Corollary \ref{cor:low_cfLimUn1} 
for $d\geq 2$.
\qed

Lemma \ref{lemma1} and Theorem  \ref{thm:low_sepLimUn1} imply the following lower bound on $\y(d,n,t)$.
\begin{theorem}\label{thm:lowerNonadapt}
Let $t$, $n$, $d$ be positive integers with $n\geq d\geq 1$. It holds
$$\y(d,n,t) \geq \max\{d\,,\,\beta\},$$
where $$
\beta\geq \cases{
\frac {\log(n+1)}2 &\hspace{-3.2truecm}if $d=1$ and $\y(d,n,t)> t/2$,\cr\cr
 {\log n\over  \log\left( \frac {et}{\y(d,n,t)}\right)}
\geq
{\log n\over \log\left( \frac {et\log(et)}{\log n}\right)}
&\hspace{-3.2truecm}if $d=1$ and $\y(d,n,t)\leq  t/2$,\cr\cr\cr
\log(n-1) &\hspace{-3.2truecm}if $d=2$ and $\y(d,n,t)> t/2$,\cr\cr
{2\log (n-1)-1\over  \log\left( \frac {et}{\y(d,n,t)}\right)}
\geq {2\log (n-1)-1\over  \log\left( \frac {et\log (et/2)}{2\log (n-1)-1}\right)}&\hspace{-3.2truecm}if $d=2$ and $\y(d,n,t)\leq t/2$,\cr\cr\cr
\!\left(\!\left\lfloor\frac{d-1}2\right\rfloor^2+\left\lfloor\frac{d-1}2\right\rfloor\!\right)\!\! \left(\frac{\log (n-\frac d2+\frac 12) }{\log\left(\frac{et (d^2-1)}{4\y(d,n,t)}\right)}-1\!\right)
\geq
\left(\!\left\lfloor\frac{d-1}2\right\rfloor^2\!+\!\left\lfloor\frac{d-1}2\right\rfloor\!\right)\!\!\left({\log \left(n-\frac d2+\frac 12\right) 
\over
\log\eta } -1\!\right) &if $d\geq 3$,\\
}$$
\hspace{.5truecm}\mbox{with $\eta= {e(d-1)^2\over  2}$ if $\
\y(d,n,t)> t/2$, and  $\eta= {2et\log({etd\over 4}) 
\over
 \log\left(n-\frac d2+\frac 12\right)-\log({etd\over 4})}$ if  $\y(d,n,t)\leq t/2$.}
\end{theorem}
\proof
Lemma \ref{lemma1} implies that $\y(d,n,t)\geq d$. The lower bounds on $\beta$ follow from the corresponding upper bounds of Theorem \ref{thm:low_sepLimUn1} on the  maximum size of a $\cup_{\leq s}$ $\dd$-separable family on the ground set $[t]$. The bounds holding for the case when $\y(d,n,t)> t/2$ and $d\leq 2$, as well as those  on the lefthand sides for the remaining cases,
are an immediate consequence of Theorem \ref{thm:low_sepLimUn1}. 
For the case when $\y(d,n,t)> t/2$ and  $d\geq 3$, the bound on the right-hand side follows from the bound on the left-hand side by simply upper bounding $\y(d,n,t)$ by $\frac t 2$. 
For $\y(d,n,t)\leq t/2$, the lower bounds on the right-hand sides are obtained as follows. Observe, that
for $\y(d,n,t)\leq t/2$, the lower bounds on the left-hand sides are 
\begin{equation}\label{rec1}
\cases{
 {\log n\over  \log\left( \frac {et}{\y(d,n,t)}\right)}
&if $d=1$,\cr\cr
{2\log (n-1)-1\over  \log\left( \frac {et}{\y(d,n,t)}\right)}
&if $d=2$,\cr\cr
\left(\!\left\lfloor\frac{d-1}2\right\rfloor^2+\left\lfloor\frac{d-1}2\right\rfloor\!\right)\!\! \left(\frac{\log (n-\frac d2+\frac 12) }{\log\left(\frac{et (d^2-1)}{4\y(d,n,t)}\right)}-1\!\right)
&if $d\geq 3$.\\}
\end{equation} 
By Lemma \ref{lemma1}, it holds  $\y(d,n,t)\geq d$, and consequently,   the above lower bounds are 
at least
\begin{equation}\label{rec2}
\cases{
{\log n\over  \log(et)}
&if $d=1$,\cr\cr
{2\log (n-1)-1\over  \log\left( \frac {et}{2}\right)}
&if $d=2$,\cr\cr
\!\left(\!\left\lfloor\frac{d-1}2\right\rfloor^2+\left\lfloor\frac{d-1}2\right\rfloor\!\right)\!\! \left(\frac{\log (n-\frac d2+\frac 12) }{\log\left(\frac{et (d^2-1)}{4d}\right)}-1\!\right)
&if $d\geq 3$.\\}
\end{equation} 
The lower bounds  on the right-hand sides for the case $\y(d,n,t)\leq t/2$ are obtained  by applying lower bounds (\ref{rec2}) to $\y(d,n,t)$ in lower bounds
(\ref{rec1}). In order to derive the bound for the case $\y(d,n,t)\leq t/2$ and $d\geq 3$, one needs also to observe that
${d^2-1 \over \left\lfloor\frac{d-1}2\right\rfloor^2+\left\lfloor\frac{d-1}2\right\rfloor} \leq 8$.
\qed

\subsection{Almost optimal  $\cup_{\leq s}$ $(p,d)$-cover free families} \label{sec:upper_pd}

The following theorem proves the existence of $\cup_{\leq s}$ $(p,d)$-cover free families with size very close to the upper bound implied 
by Theorem \ref{thm:low_p_d_cf_low}. %The  proof of this theorem is in the Appendix.

\begin{theorem}\label{thm:p_d_cf_upp}
Let $d$ and $p$ be positive integers and let $s$ and $t$ be integers such that $ s\leq t$. There exists a $\cup_{\leq s}$ $(p,d)$-cover free family
on the ground set $[t]$ with size 
$$n\geq
\cases{
  \frac1e(p+d) 2^{\left({p\over d(d+p)}\left(s-d\log\left(\frac{e(d+p)}p\right)-\frac dp\right)\right)}&if $t<2s$,\cr\cr
\frac 1e(p+d) 2^{\left({p\over d(d+p)}\left(s\log\left(\frac{et}s\right)-d\log\left(\frac{e(d+p)}p\right)-\frac dp\right)\right)} &if $t\geq2s$.\\
}$$
\end{theorem}
\proof
 We will prove the theorem by the probabilistic method.
In the following, we will conveniently represent a family $\F$ of $n$ subsets of $[t]$
by the  $t\times n$ binary matrix having as columns the characteristic vectors $\cc_1,\ldots,\cc_n$ of the subsets belonging to $\F$, i.e., for each $i=1,\ldots,t$ and
$j=1,\ldots,n$, the matrix has entry $(i,j)$ set to 1 if and only if the member of $\F$ associated with the $j$-th column contains $i$.
The number of 1-entries of a column $\cc$ will be called the {\em weight} of $\cc$. 
Given $m$ columns  $\cc_{j_1},\ldots,\cc_{j_m}$, we will denote by $\cc_{j_1}\vee \ldots\vee\cc_{j_m}$ the Boolean $OR$ of columns $\cc_{j_1},\ldots,\cc_{j_m}$.

Let us consider a $t\times n$ random binary matrix $\M$ where each entry is 0 with probability $\yy$ and 1 with probability
$1-\yy$, with $\yy=\left(1-\left(\frac {s}{et}\right)^{s(\frac pd+1)}\right)^{\frac 1d}$. In order for $\M$ to represent a $\cup_{\leq s}$ $(p,d)$-cover free family, it must hold that for any choice
of $d$ columns $\cc_{j_1}, \ldots, \cc_{j_d}$ the following two events $E_1$ and $E_{2}$ occur.
\begin{itemize}
\item[$E_1$:] The weight of $\cc_{j_{1}}\vee\ldots \vee \cc_{j_{d}}$  is at most $s$, i.e., there is a number $a$ of rows,
 $a\leq s$,  such that in correspondence of each of these $a$ rows at least  one of $\cc_{j_1}, \ldots, \cc_{j_d}$ has an entry equal to 1, whereas  in correspondence of the remaining $t-a$ rows, all entries of $\cc_{j_1}, \ldots, \cc_{j_d}$ are equal to 0. 
 \item[$E_2$:]  For any choice
of $p$ other columns $\cc_{k_1}, \ldots, \cc_{k_p}$,  the column $\cc_{j_1}\vee \ldots \vee \cc_{j_d}$ does not cover the  column $\cc_{k_{1}}\vee\ldots\vee \cc_{k_{p}}$, i.e., there exists a row index $i$ such that at least one of $\cc_{k_1},\ldots,\cc_{k_p}$ has the $i$-th entry equal to 1 whereas all
columns $\cc_{j_{1}},\ldots, \cc_{j_{d}}$ have the $i$-th entry equal to 0.
\end{itemize}
We say that a set of  $d$ columns $\{\cc_{j_{1}}\vee\ldots \vee \cc_{j_{d}}\}$    is {\em good} if both events $E_1$ and $E_2$  occur.  
We will prove that  the probability that $\M$ contains  a set of $d$ columns which is not good is smaller than
1, thus proving that $\M$ has a positive probability of representing a $\cup_{\leq s}$ $(p,d)$-cover free family.

For a given set of $d$ columns  $\cc_{j_1}, \ldots, \cc_{j_d}$ of $\M$, we want to estimate  probability

%\begin{equation}
%Pr\{ \cc_{j_1}, \ldots, \cc_{j_p} \mbox{ is bad }\} =1- Pr\{ \cc_{j_1}, \ldots, \cc_{j_p} \mbox{ is not bad }\}.
%\end{equation}

\begin{eqnarray}
Pr\{ \{\cc_{j_1}, \ldots, \cc_{j_d} \}\mbox{ is good}\}
&=&Pr\{E_1 \cap E_2\}
= Pr\{E_2 | E_1\} Pr\{E_1\} \label{eq:pgood}
\cr\cr
&=& \left(1-Pr\{\overline {E_2} | E_1\}\right) Pr\{E_1\}.
\end{eqnarray}

Let us estimate the probability $Pr\{\overline {E_2} | E_1\}$. Notice that event $E_1$ implies that there are at most $s$ entries equal to 1 in  $\cc_{j_1}\vee \ldots \vee \cc_{j_d}$. Let $0\leq a\leq s$ be an integer and let $i_1,\ldots, i_a$  be $a$ row  indices of $\M$. We denote by $E_{i_1,\ldots,i_a}$ the event that the vector  $\cc_{j_{1}}\vee \ldots \vee \cc_{j_{d}}$ has all entries with indices in $\{i_1,\ldots, i_a\}$ equal to 1 and all other entries equal to 0.
For the given set of row indices $\{i_1,\ldots, i_a\}$, let us estimate the probability $Pr\{\overline {E_2} \cap E_{i_1,\ldots,i_a} | E_1\}$.
\begin{eqnarray}
&&Pr\{\overline{E_2} \cap E_{i_1,\ldots,i_a} | E_1\}   \cr\cr
&=& Pr\{\overline {E_2} | E_{i_1,\ldots,i_a} \cap E_1\} \cdot Pr\{E_{i_1,\ldots,i_a} | E_1\} \cr\cr
&=& Pr \left\{\exists\, 
\cc_{k_1},\ldots, \cc_{k_p}\not\in  \{\cc_{j_1},\ldots,\cc_{j_d}\} \mbox{ such that }
 \cc_{k_1}\vee\ldots\vee\cc_{k_p} \mbox{ is covered by }\right.\cr
&&\,\,\,\,\,\,\,\,\,\,\left. \cc_{j_1}\vee\ldots\vee\cc_{j_d}| E_{i_1,\ldots,i_a} \cap E_1\right\} \cdot Pr\{E_{i_1,\ldots,i_a} | E_1\} \cr\cr
&=& Pr \left\{\exists\, 
\cc_{k_1},\ldots, \cc_{k_p}\not\in  \{\cc_{j_1},\ldots,\cc_{j_d}\}\mbox{ such that }
 (\cc_{k_1}\vee\ldots\vee\cc_{k_p})(i)=0,\right.\cr
&&\,\,\,\,\,\,\,\,\,\, \left.\mbox{ for all  } i\in[t]\setminus\{i_1,\ldots,i_a\} \right\}
 \cdot Pr\{E_{i_1,\ldots,i_a} | E_1\} \label{eq:prob1}\\\cr
&\leq&\left[ {n-d\choose p} \yy^{p(t-a)}\right]\cdot \left[(1-\yy^d)^a\yy^{d(t-a)}\right]. \label{eq:prob2}
\end{eqnarray}
The second term in (\ref{eq:prob2}) has been obtained by observing that $Pr\{E_{i_1,\ldots,i_a} | E_1\}=Pr\{E_{i_1,\ldots,i_a} \}$.

Notice that for $\{i_1,\ldots.i_a\}\neq \{i'_1,\ldots,i'_{a'}\}$, with  $0\leq a\leq s$ and $0\leq a'\leq s$, it is $E_{i_1,\ldots.i_a}\cap E_{i'_1,\ldots,i'_{a'}}=\emptyset$.
By the law of total
probability and upper bound (\ref{eq:prob2}), we have that
\begin{eqnarray}
Pr\{\overline {E_2} | E_1\}&=&\sum_{a=0}^s\sum_{(i_1,\ldots,i_a)\in[t]_a} Pr\{\overline {E_2} \cap E_{i_1,\ldots,i_a} | E_1\}\cr\cr
&\leq&\sum_{a=0}^s{t\choose a} \left[ {n-d \choose p}\yy^{(p+d)(t-a)}(1-\yy^d)^{a}\right]\cr\cr
&\leq&  {n-d\choose p} (1-\yy^d) \yy^{(p+d) (t-s)}\sum_{a=0}^s{t\choose a}.\,\label{eq:pbad1}
%\cr\cr
%&\leq&  {n-d\choose p} (1-\yy^d) \yy^{(p+d) (t-s)}2^{s\log(et/s)},\,\label{eq:pbad1}
\end{eqnarray}
%where the last inequality follows from (\ref{eq:leftLowAdap2}) which implies 
%that $\sum_{a=0}^s{t\choose a}\leq 2^{s\log \frac {et}s}$, for $s\leq t/2$.

By upper bound (\ref{eq:pbad1}) and by (\ref{eq:pgood}), we have that
\begin{equation}
Pr\{ \{\cc_{j_1}, \ldots, \cc_{j_d} \}\mbox{ is good}\}
\geq \left(1-{n-d\choose p} (1-\yy^d) \yy^{(p+d)(t-s)}\sum_{a=0}^s{t\choose a}\right)\cdot Pr\left\{E_1\right\}.
\end{equation}
Now let us estimate $Pr\left\{E_1\right\}$, that is the probability that  $\cc_{j_1}\vee\ldots\vee\cc_{j_d}$ 
has weight at most $s$. For a fixed row index $i$, the probability that $\cc_{j_1}\vee\ldots\vee\cc_{j_d}$ has the $i$-th entry equal to 1 is $(1-\yy^d)$. For $i=1,\ldots, t$, let $X_i$ be the Bernoulli random variable which is 1 if and only if at least one of $\cc_{j_1},\ldots,\cc_{j_d}$ has the $i$-th entry equal to 1. Therefore, the  random variable $\sum_{i=1}^t X_i$ has a binomial distribution with probability of success equal to
$(1-\yy^d)$. By Markov inequality, the probability that $\sum_{i=1}^t X_i>s$ is at most $\frac{E[\sum_{i=1}^t X_i]}{s+1}= \frac{t(1-\yy^d)}{s+1}$, thus implying that $Pr\left\{E_1\right\}\geq \left(1-\frac{t(1-\yy^d)}{s+1}\right)$. It follows that

\begin{eqnarray}
&&Pr\{ \{ \cc_{j_1}, \ldots, \cc_{j_d} \}\mbox{ is good}\}
\cr\cr
&\geq& \left(1- {n-d\choose p} (1-\yy^d) \yy^{(p+d)(t-s)}\sum_{a=0}^s{t\choose a}\right) \left(1-\frac{t(1-\yy^d)}{s+1}\right)\cr\cr
&=&  1-\frac{t(1-\yy^d)}{s+1}- \left(1-\frac  {t(1-\yy^d)}{s+1}\right){n-d\choose p} (1-\yy^d) \yy^{(p+d)(t-s)}\sum_{a=0}^s{t\choose a} \cr\cr
&\geq&1 -\frac{t(1-\yy^d)}{s+1}- {n-d\choose p} (1-\yy^d)\sum_{a=0}^s{t\choose a}.
\label{eq:pgood2}
\end{eqnarray}

Now we are ready to estimate the probability that $\M$ does not represent a $\cup_{\leq s}$ $(p,d)$-cover free family.
Inequality (\ref{eq:pgood2}) allows to upper bound the probability that a given set of $d$ columns is not good. Therefore, we have that
\begin{eqnarray}
&&Pr\left\{\M \mbox{ does not represent a  $\cup_{\leq s}$ $(p,d)$-cover free  family}\right\}\cr\cr
&=&Pr\left\{\mbox{there exists a set }  \{\cc_{j_1}, \ldots, \cc_{j_d} \} \mbox{ which is not good } \right\}\cr\cr
%&\leq&   
%{n\choose d} \left(\frac{t(1-\yy^d)}{s+1}+2^{s\log(et/s)} {n-d\choose p}  (1-\yy^d) \yy^{(p+d)(t-s)}\left(1-\frac{t(1-\yy^d)}{s+1}\right)\right)\cr\cr
&\leq&   
{n\choose d} \left(\frac{t(1-\yy^d)}{s+1}+ {n-d\choose p} (1-\yy^d)\sum_{a=0}^s{t\choose a}%y^{(p+d)(t-s)} 
\right). \label{eq:pgood3}
\end{eqnarray}

By setting $\yy=\left(1-\left(\frac {s}{et}\right)^{s(\frac pd+1)}\right)^{\frac 1d}$ in (\ref{eq:pgood3}) we obtain that

\begin{eqnarray}
&&Pr\left\{\M \mbox{ does not represent a  $\cup_{\leq s}$ $(p,d)$-cover free  family}\right\}\cr\cr
&\leq&  {n\choose d} \frac{t}{s+1} \left(\frac{s}{et}\right)^{s(\frac pd+1)}+
{n\choose d}{n-d\choose p} \left(\frac{s}{et}\right)^{s(\frac pd+1)}\sum_{a=0}^s{t\choose a}\cr\cr
&<& 2
{n\choose d}{n-d\choose p} \left(\frac{s}{et}\right)^{s(\frac pd+1)}\sum_{a=0}^s{t\choose a}\cr\cr
&=&2{n\choose d+p}{d+p\choose p} \left(\frac{s}{et}\right)^{s(\frac pd+1)}\sum_{a=0}^s{t\choose a}.\label{eq:bP}
%&\leq&  2
%{n\choose d}{n-d\choose p} 2^{s\log(et/s)} \left(\frac{s}{et}\right)^{s(\frac pd+1)}.%\cr\cr
%&=& 2\cdot 2^{s\log(et/s)}2^{-s\log(et/s)(\frac pd+1)}{n\choose d}{n-d\choose p}\cr\cr
%&=&2\cdot 2^{-\frac {sp}d\log(et/s)}{n\choose d}{n-d\choose p}. \label{eq:bP}
 \end{eqnarray}

Let  $P$ denote  $Pr\left\{\M \mbox{ does not represent a  $\cup_{\leq s}$ $(p,d)$-cover free  family}\right\}$. By (\ref{eq:bP}), we have that 
\begin{equation}\label{eq:P}
P< 2
{n\choose d+p}{d+p\choose p} \left(\frac{s}{et}\right)^{s(\frac pd+1)}\sum_{a=0}^s{t\choose a}.
\end{equation}
In order  for a $\cup_{\leq s}$ $(p,d)$-cover free family of size $n$ on the ground set $[t]$ to exist, 
it is sufficient that  $P<1$.  

We first consider the case  $t\geq 2s$ and then  the case $t<2s$.

For $t\geq 2s$, inequality (\ref{eq:leftLowAdap2}) implies 
that $\sum_{a=0}^s{t\choose a}\leq 2^{s\log \frac {et}s}$,  and consequently, by (\ref{eq:P}) we have that
\begin{equation}\label{eq:Pc1}
P<2
{n\choose d+p}{d+p\choose p} 2^{-s\log \frac {et}s(\frac pd+1)}2^{s\log \frac {et}s}.
\end{equation}
By (\ref{eq:Pc1}), one has that $P<1$ holds if
$${sp\over d}\log(et/s)>\log\left(2{n\choose d+p}{d+p\choose p}\right).$$
Therefore, one has that $P<1$ if 
\begin{equation}
s%&>&{d\over p+d}\cdot{\log\left(2{n\choose d}{n-d\choose p}\right)\over  \log \left(\frac{et}{s}\right)-{d\over p+d}\log\left(\frac {et}s\right)}\cr\cr\cr
%&=&{d\over p+d}\cdot{\log\left(2{n\choose d+p}{d+p\choose p}\right)\over  \frac {p}{p+d}\log \left(\frac{et}{s}\right)}\cr\cr\cr
\geq {d\over p}\cdot{\log\left(2{n\choose d+p}{d+p\choose p}\right)\over  \log \left(\frac{et}{s}\right)}. \label{eq:inP1}
\end{equation}

 By  the upper bound in (\ref{eq:binomial_upp}),  we can limit from above the binomial
 coefficients in the right-hand side  of  (\ref{eq:inP1}), and obtain that there exists a   $\cup_{\leq s}$ $(p,d)$-cover free family of size $n$ on the ground set $[t]$ if 
 $$s\geq {{d(d+p)\over p}\log\left({en\over d+p}\right) + {d} \log\left({e(d+p)\over p}\right)+{d\over p}\over  \log \left(\frac{et}{s}\right)},$$
 which is satisfied for any
 $n\leq  \frac1e(p+d) 2^{\left({p\over d(d+p)}\left(s\log\left(\frac{et}s\right)-d\log\left(\frac{e(d+p)}p\right)-\frac dp\right)\right)}$. Therefore, 
 we have that,  for $t\geq 2s$, there exists a   $\cup_{\leq s}$ $(p,d)$-cover free family of size $n$ on the ground set $[t]$ 
 that satisfies the  second bound in the statement of the theorem.
 
Now, let us consider the case $t<2s$. In this case, we observe that  $ \left(\frac{s}{et}\right)^{s(\frac pd+1)}$  decreases with $s$ and therefore, 
we can limit it from above by  $\left(\frac{1}{2e}\right)^{\frac t2(\frac pd+1)}$ in the right-hand side of
(\ref{eq:P}). Moreover, we  upper bound $\sum_{a=0}^s{t\choose a}$ by $2^t$. Consequently, one has 
\begin{eqnarray}
P&<& 2
{n\choose d+p}{d+p\choose p} 2^t \left(\frac{1}{2e}\right)^{\frac t2(\frac pd+1)}\cr\cr
&<&2
{n\choose d+p}{d+p\choose p}2^t \left(\frac{1}{2}\right)^{t(\frac pd+1)}\cr\cr
&=&2 {n\choose d+p}{d+p\choose p} \left(\frac{1}{2}\right)^{t(\frac pd)}\cr\cr
&\leq& 2{n\choose d+p}{d+p\choose p}\left(\frac{1}{2}\right)^{s(\frac pd)},\label{eq:P2}
\end{eqnarray}
where the last inequality follows from $s$ being smaller than or equal to $t$.

Therefore, one has that $P<1$ if 
\begin{equation}
s
\geq {d\over p}\cdot{\log\left(2{n\choose d+p}{d+p\choose p}\right)}. \label{eq:inP}
\end{equation}

 By  the upper bound in (\ref{eq:binomial_upp}),  we can limit from above the binomial
 coefficients in the right-hand side  of  (\ref{eq:inP}), and obtain that there exists a   $\cup_{\leq s}$ $(p,d)$-cover free family of size $n$ on the ground set $[t]$ if 
 $$s\geq  {{d(d+p)\over p}\log\left({en\over d+p}\right) + {d} \log\left({e(d+p)\over p}\right)+{d\over p}},$$
 which is satisfied for any
 $n\leq  \frac1e(p+d) 2^{\left({p\over d(d+p)}\left(s-d\log\left(\frac{e(d+p)}p\right)-\frac dp\right)\right)}$.
 It follows  that,  for $t< 2s$, there exists a   $\cup_{\leq s}$ $(p,d)$-cover free family of size $n$ on the ground set $[t]$ 
 that satisfies the first bound in the statement of the theorem.
\qed

  In the following, we compare the lower bounds  of Theorem \ref{thm:p_d_cf_upp} with the upper bounds of  Theorem \ref{thm:low_p_d_cf_low}. In fact, we will estimate the gap between the upper and lower bounds on $\log(n_{cf}(p,d,\cup_{\leq s},t))$, thus showing that this gap is not larger than that  existing between the best upper and lower bounds on the logarithm of the maximum size of classical $(p,d)$-cover free families.
For the case $t\geq 2s$, Theorem \ref{thm:p_d_cf_upp} implies an 
$\Omega\left(\frac {sp}{d(p+d)}\log \left(\frac{et}s\right)\right)$ lower bound
 on $\log(n_{cf}(p,d,\cup_{\leq s},t))$.
  Theorem \ref{thm:low_p_d_cf_low} implies that  	       $\log(n_{cf}(p,d,\cup_{\leq s},t))$ is upper bounded by 
 $O\left(\frac sd\log\left(\frac{et}{s}\right)\right)$   for $d<2p$, and by 
 $O\left(\frac {sp}{d^2}\log\left(\frac{etd^2}{4 ps}\right)\right)$  for $d\geq 2p$. Therefore, for $d<2p$, the gap between the upper and lower bounds on $\log(n_{cf}(p,d,\cup_{\leq s},t))$ is $O\left({\frac sd\log\left(\frac{et}{s}\right)\over \frac {sp}{d(p+d)}\log \left(\frac{et}s\right)}\right)= 
O\left({\frac sd\log\left(\frac{et}{s}\right)\over \frac s{d}\log \left(\frac{et}s\right)}\right)= O(1)$. For $d\geq 2p$, 
the gap is limited from above by 
$$O\left({\frac {sp}{d^2}\log\left(\frac{etd^2}{4ps}\right)\over \frac {sp}{d(p+d)}\log \left(\frac{et}s\right)}
\right)=O\left({\frac {sp}{d^2}\log\left(\frac{etd^2}{4ps}\right)\over \frac {sp}{d^2}\log \left(\frac{et}s\right)}
\right)=O\left(1+{\log\left(\frac{d^2}{4p}\right)\over \log\left(\frac{et}{s}\right)}
\right).$$
Interestingly,  the above bound decreases as the ratio  between the size of the ground set $t$ and the bound on the number of elements in the union of any $d$ members of the family increases.
 If we set $s=t$ in the above bound, we obtain the same asymptotic gap existing  between the best upper and lower bounds on the logarithm of the maximum size of classical $(p,d)$-cover free families.

 For the case $t<2s$,  Theorem \ref{thm:p_d_cf_upp} implies that  $\log(n_{cf}(p,d,\cup_{\leq s},t))$  is $\Omega\left({sp\over d(d+p)}\right)$. 
   Theorem \ref{thm:low_p_d_cf_low} implies that   $\log(n_{cf}(p,d,\cup_{\leq s},t))$ is upper bounded by 
 $O\left(\frac sd\right)$   for $d<2p$, and by 
 $O\left(\frac {sp}{d^2}\log\left(\frac{etd^2}{4 ps}\right)\right)=O\left(\frac {sp}{d^2}\log\left(\frac{d^2}{ p}\right)\right)$  for $d\geq 2p$. For $d<2p$, one has   $\Omega\left({sp\over d(d+p)}\right)= \Omega\left({s\over d}\right)$, and consequently, the lower bound on $\log(n_{cf}(p,d,\cup_{\leq s},t))$ asymptotically matches the upper bound. For $d\geq 2p$, one has   $\Omega\left({sp\over d(d+p)}\right)= \Omega\left({sp\over d^2}\right)$ and the ratio between the upper and lower bounds on
 $\log(n_{cf}(p,d,\cup_{\leq s},t))$ is 
 $$O\left({\frac {sp}{d^2}\log\left(\frac{d^2}{p}\right)\over \frac {sp}{d^2}}
\right)=O\left(\log\left(\frac{d^2}{p}\right) 
\right),$$ which is the same gap existing between the best upper and lower bounds on the logarithm of the maximum
size of classical  $(p,d)$-cover free families.

 By setting $p=1$ in the bound of Theorem \ref{thm:p_d_cf_upp},
we obtain the following lower bound on the maximum size of  $\cup_{\leq s}$ $d$-cover free families
on the ground set $[t]$.
  \begin{theorem}\label{thm:1_d_cf_upp}
Let $d$  be a positive integer and let $s$ and $t$ be integers such that $ s\leq t$. There exists a $\cup_{\leq s}$ $d$-cover free family
on the ground set $[t]$ with size 
$$n\geq
\cases{
  \frac1e(d+1) 2^{\left({1\over d(d+1)}\left(s-d\log({e(d+1)})- d\right)\right)}&if $ t<2s$,\cr\cr
\frac 1e(d+1) 2^{\left({1\over d(d+1)}\left(s\log\left(\frac{et}s\right)-d\log\left({e(d+1)}\right)-d\right)\right)} &if $t\geq2s$.\\
}$$
\end{theorem}

The above theorem implies the following upper bound on the number of ``yes" responses admitted by a non-adaptive group testing algorithm 
that uses at most $t$ tests.
\begin{theorem}\label{thm:upper_nonA1}
Let $t$, $n$, $d$ be positive integers with $d\geq 1$ and $n\geq d$. There exists a non-adaptive group testing strategy $\A$ for which $\y_\A(d,n,t)$ is at most
   $$
    \cases{
 {{d(d+1)}\log\left({en\over d+1}\right) + {d} \log\left({e(d+1)}\right)+{d}}  & if $ t<2\y_\A(d,n,t)$,\cr\cr
  %{d(d+1)\over  \log \left(\frac{et}{\y_\A(d,n,t)}\right)}\left(\log\left({en\over d+1}\right) + \frac {\log\left({e(d+1)+1}\right)}{d+1} \right)
{d(d+1)\over  \log \left(\frac{et}{\y_\A(d,n,t)}\right)}\left(\log\left({en\over d+1}\right) + \frac {\log({e(d+1))+1}}{d+1} \right) \leq 
{d(d+1)\over \log\mu}\left(\log\left({en\over d+1}\right) + \frac {\log({e(d+1))+1}}{d+1} \right) 
&if $t\geq 2\y_\A(d,n,t)$,\\
    }
 $$
 where $\mu = \log \left(\frac{et\log (2e)}  {d(d+1)\left(\log\left(\frac {en}{d+1}\right)+\log (2\sqrt e)\right ) }\right ) $.
 %{d^2\log\left({n\over d}\right) )\over  \log \left(\frac{et}{\y_\A(d,n,t)}\right)}\right)=
%O\left({d^2\log\left({n\over d}\right) )\over  \log \left(\frac{et \log \left(\frac{et}{\y_\A(d,n,t)}\right)}{d^2\log\left({n\over d}\right) )}\right)}\right).$$ 
\end{theorem}
\proof
The upper bound for $t <2\y_\A(d,n,t)$ follows immediately from the lower bound in Theorem \ref{thm:1_d_cf_upp}.
For $t \geq 2\y_\A(d,n,t)$, Theorem \ref{thm:1_d_cf_upp} implies 
\begin{equation}\label{eq:4_8}\y_\A(d,n,t) \leq {d(d+1)\over  \log \left(\frac{et}{\y_\A(d,n,t)}\right)}\left(\log\left({en\over d+1}\right) + \frac {\log({e(d+1))+1}}{d+1} \right).\end{equation}
Since  $ \frac {\log({e(d+1))+1}}{d+1}$ decreases with $d$, we upper bound  it  by $\log (2\sqrt e)$  in (\ref{eq:4_8}) and obtain 
\begin{equation}\label{eq:recUp}
\y_\A(d,n,t) \leq {d(d+1)\over  \log \left(\frac{et}{\y_\A(d,n,t)}\right)}\left(\log\left({en\over d+1}\right) + \log (2\sqrt e)\right).\end{equation}
In order to derive an upper bound on $\y_\A(d,n,t)$, expressed in terms of $d$, $n$, and $t$ only, 
we first exploit upper bound (\ref{eq:recUp}) to limit from above $\y_\A(d,n,t)$ in upper bound (\ref{eq:4_8}), thus obtaining
\begin{equation}\label{eq:up}
\y_\A(d,n,t)\leq
 {d(d+1) \left(\log\left(\frac {en}{d+1}\right) + \frac {\log(e(d+1))+1}{d+1} \right)
\over  
\log \left(
\frac{et\log \left(\frac{et}{\y_\A(d,n,t)}\right)}  {d(d+1)\left(\log\left(\frac {en}{d+1}\right)+\log (2\sqrt e)\right )}   \right ) }.
\end{equation}
Then, we upper bound $\y_\A(d,n,t)$ in
 (\ref{eq:up}) by $t/2$ thus obtaining $\log \left(\frac{et}{\y_\A(d,n,t)}\right)\geq \log (2e)$, and consequently, the upper bound that appears on the right-hand side of case $t \geq 2\y_\A(d,n,t)$. \qed

The result of Theorem \ref{thm:p_d_cf_upp} will be exploited in Section \ref{sec:two_stage} to prove the existence of a trivial two-stage algorithm that admits the same number of positive responses of the best adaptive procedures.  

\subsection{An almost optimal explicit non-adaptive algorithm}\label{sec:explicit}
In this section  we present another non-adaptive  algorithm that  gets very close to the lower bound of Theorem \ref{thm:lowerNonadapt}.

We remark that this result  translates into a lower bound on the size of $\cup_{\leq s}$ $d$-cover free families which is very close to the upper bound of Corollary \ref{cor:low_cfLimUn1}. The underlying combinatorial structures of the algorithm consist of  families in which any two members share at most a certain number  $\lambda$ of   elements.
The following simple lemma will be used in the analysis of both algorithms.
% It can be easily proved by induction.
\begin{lemma} \label{lemma2}
Let  $d$ and $\lambda$ be two positive integers  and let $\F$ be a  family of sets with $|\F|\geq d$ and  such that any two members $F_1,F_2\in \F$ intersect in at most $ \lambda$ elements.  Then, for any $d$ members $F_1,\ldots,F_d$ of $\F$, it holds $|\bigcup_{i=1}^d F_i |\geq  \sum_{i=1}^d |F_{i}|-\frac12d(d-1)\lambda$.
\end{lemma}
\proof
Observe that 
\begin{equation}\label{eq:lemma2_1}
\big |\bigcup_{i=1}^d F_i \big |\geq  \big |\bigcup_{i=1}^d (F_i\setminus \bigcup_{j=1}^{i-1}\big ( F_i\cap F_j))\big |.\end{equation}
Since for $i\neq \ell$, it holds  $\left(F_i\setminus \bigcup_{j=1}^{i-1} (F_i\cap F_j)\right)\cap \left(F_\ell\setminus \bigcup_{j=1}^{\ell-1} (F_\ell\cap F_j)\right)=\emptyset$,
one has that the right-hand side of (\ref{eq:lemma2_1}) is equal to
\begin{equation}\label{eq:lemma2_2} 
 \sum_{i=1}^d \big |F_i\setminus \bigcup_{j=1}^{i-1} ( F_i\cap F_j )\big |\end{equation}
Notice that for any two sets $A$ and $B$, one has that  $|A\setminus B|\geq |A|-|B|$, with equality holding if and only if
$B\subseteq A$. Therefore, it holds $|F_i\setminus \bigcup_{j=1}^{i-1} ( F_i\cap F_j )|\geq
|F_i|-\big| \bigcup_{j=1}^{i-1} ( F_i\cap F_j )\big| $, and consequently,  expression  (\ref{eq:lemma2_2}) is larger than or equal to
$$  \sum_{i=1}^d \big |F_i\big |- \big |\bigcup_{j=1}^{i-1}\big ( F_i\cap F_j)\big |= \sum_{i=1}^d \big |F_i\big |- \sum_{i=1}^d  \big |\bigcup_{j=1}^{i-1}( F_i\cap F_j)\big |
\geq \sum_{i=1}^d \big |F_i\big |- \sum_{i=1}^d  \sum_{j=1}^{i-1}\big  |F_i\cap F_j|.$$
Since $\sum_{i=1}^d  \sum_{j=1}^{i-1} |F_i\cap F_j|\leq \sum_{i=1}^d  \sum_{j=1}^{i-1} \lambda={d\choose 2} \lambda$, the lemma follows.
\qed

An interesting feature of the construction presented in this section is that it is an explicit construction. It is based on a breakthrough result by Porat and Rothschild \cite{porat} which provides the first deterministic explicit construction of error correcting codes meeting the Gilbert-Varshamov bound. In fact, the result in \cite{porat} provides a construction for  $[m,k,\delta m]_q$-linear codes. 
We recall that an $[m,k,\delta m]_q$-linear code is a $q$-ary code over the alphabet $\mathbb{F}_q$ with length $m$, size $n=q^k$ and  Hamming distance equal to  $\delta m$. In the following, we denote by $H_q( p )$ the $q$-ary entropy function
$$H_q( p )= p \log_q \frac{q-1}{p}+(1-p)\log _q \frac {1}{1-p},$$
which, with respect to the Hamming distance over $q$-ary alphabets  plays a role analogue to that played by  binary entropy with respect to the binary alphabet.
Porat and Rothschild proved the following 
\begin{theorem}\label{thm:prLC}\cite{porat}
Let $q$ be a prime power, $m$ and $k$ positive integers, and $\delta \in [0,1]$. If $k\leq (1-H_q(\delta))m$, then it is possible to construct an $[m,k,\delta m]_q$-linear code  in time $\Theta(mq^k)$.   
\end{theorem}

 In \cite{porat}, Porat and Rothschild show how to construct an $(n,r)$-strongly selective family \cite{sel2} from a linear code with properly chosen parameters and then exploit the above mentioned theorem to construct in time $ \Theta(r n\ln n)$ a linear code that can be reduced to an
  $(n,r)$-{\em strongly selective} family of size $\Theta(r^2\ln n)$.  We just mention that an $(n,r)$-strongly selective family is a combinatorial structure which is essentially equivalent to an $(r-1)$-cover free family.
  The following theorem rephrases the result in \cite{porat} in terms of cover free families.

 \begin{theorem}\label{thm:pr_reducition}
If there exists an $[m,k,\delta m]_q$-linear code then it is
possible to construct an $m$-uniform $(\lceil \frac{1}{ 1-\delta}\rceil-1)$-cover free family of size $n=q^k$
on the ground set $[mq]$, with the property that any two members of the family intersect in at most $m-\delta m$ elements.
\end{theorem}

\proof
Given an $[m,k,\delta m ]_q$-linear code $\C=\{\cc_1,\ldots,\cc_n\}$, let us define the family $\F$ as 
$\F=\{F(\cc_1),\ldots, F(\cc_n)\}$, where $F(\cc_j)=\{f(i,a): (i,a)\in[m]\times  [q], \,\cc_j[i]=a\} $, with $f$ being an injection 
from $[m]\times  [q] $ to $[m q]$.
It is immediate to see that $\F$ is $m$-uniform in that for each index $i\in [m]$  there is a unique pair $(i,a)\in[m]\times[q]$ such that  $\cc_j[i]=a$.
Moreover, any two members of $\F$  intersect in at most $m-\delta m$ elements. Indeed, for any two distinct words $\cc_j,\cc_\ell\in \C$ there are at least $\delta m$ indices $i\in[m]$
such that $\cc_\ell(i)\neq \cc_j(i)$. This implies that there are at least $\delta m$ pairs $(i,a)\in[m]\times  [q] $ such that $f(i,a)\in F(\cc_j)$ and $f(i,a)\not\in F(\cc_\ell)$, and consequently,  $F(\cc_j)$ and $F(\cc_\ell)$ share at most $m-\delta m$ elements.
It follows that the union of any $\lceil \frac{1}{ 1-\delta }\rceil-1$ members of $\F$ shares at most $(\lceil \frac{1}{ 1-\delta }\rceil-1) (m-\delta m) \leq m-1$ elements with any other member of the family, implying that $\F$ is $(\lceil \frac{1}{ 1-\delta }\rceil-1)$-cover free.
\qed

\begin{theorem}\label{thm:explicit_construction}
Let $t$, $n$, $d$ be positive integers with $n\geq d\geq 1$. There exists a non-adaptive group testing strategy $\A$ for which
$$\y_\A(d,n,t) =\cases{\Theta(d^2\ln n)& if $\y_\A(d,n,t)\geq \frac{(t+1)d}{2(d+1)}$,\cr\cr  
\Theta\left({d^2\ln n\over{\ln(\frac t {\y_\A(d,n,t)}})} \right)=O\left( {d^2\ln n
\over  
\ln \left(
\frac{t}{d^2\ln n}\right)}\right)& if $\y_\A(d,n,t)< \frac{(t+1)d}{2(d+1)}$\\}. $$
The underlying family can be constructed in time $\Theta (dn\ln n)$ if $\y_\A(d,n,t)\geq \frac{(t+1)d}{2(d+1)}$, and in time $\Theta\left({dn\ln n\over{\ln(\frac t {\y_\A(d,n,t)})} }\right)=
O\left({dn\ln n\over{\ln(\frac t {d^2\ln n}} )}\right)$ otherwise.
\end{theorem}
 
 \proof
 For $\y_\A(d,n,t)\geq\frac{(t+1)d}{2(d+1)}$, the stated bound follows from Theorem  1 of \cite{porat} which implies that there exists a non-adaptive group testing algorithm that uses
 $t=\Theta(d^2\ln n)$ and is such that the underlying  family can be constructed in time $\Theta(dn\ln n)$. Since in the case we are considering it is $\y_\A(d,n,t)=\Theta( t)$, we have $\y_\A(d,n,t)=\Theta(d^2\ln n)$. 
  
  Let us consider the case when  $\y_\A(d,n,t)< \frac{(t+1)d}{2(d+1)}$.
 By Theorem \ref{thm:prLC} it is possible to construct an  $[m,k,\delta m]_q$ linear code  in time $\Theta(mq^k)$, where $q$ is a prime power, $m$ a positive integer, $\delta \in [0,1]$ and $k= (1-H_q(\delta))m$.
 Theorem \ref{thm:pr_reducition} then implies that such a code can be transformed into an $m$-uniform $(\lceil \frac{1}{ 1-\delta}\rceil-1)$-cover free family $\F$ of size $n=q^k$ on the ground set $[mq]$. Let us set $\delta = \frac d{d+1}$, and let $q\geq 2d+2$.
It holds
\begin{eqnarray}
1-H_q(\delta)&=&
1-\left[\frac d{d+1}\log_q\left(\frac {(d+1)(q-1)}{d}\right) + \frac 1 {d+1}\log_q(d+1)\right]\cr\cr
&=&\frac 1{(d+1)\ln q}\left[(d+1)\ln q-d\ln\left(\frac {(d+1)(q-1)}{d}\right) - \ln(d+1)\right]\cr\cr
&=&\frac 1{(d+1)\ln q}\left[d\ln q+\ln q-d\ln\left(\frac {d+1}{d}\right) -d\ln(q-1)- \ln(d+1)\right]\cr\cr
&=&\frac 1{(d+1)\ln q}\left[d\ln \left(\frac q{q-1}\right)-d\ln \left(\frac {d+1}d\right)   +\ln\left(\frac q{d+1}\right)\right]. \label{eq:last}
\end{eqnarray}
We can exploit the well known relation $\ln\frac z{z-1}=\frac 1 z+o(\frac 1 z)$, to estimate (\ref{eq:last}). Therefore, we get
\begin{equation}
1-H_q(\delta)=
\frac 1{(d+1)\ln q}\left[ \frac dq- \frac d{d+1}+ \ln\left(\frac q{d+1}\right)\right]+o\left(\frac 1{(d+1)\ln q}\right).
\end{equation}
We will prove that  $$ c \cdot \ln\left( \frac q{d+1}\right)\leq \left[ \frac dq- \frac d{d+1}+ \ln\left(\frac q{d+1}\right)\right]< \ln\left(\frac q{d+1}\right),$$ for any constant $c\leq 1/6$.
Indeed, we are assuming $q\geq 2d+2$ and therefore,
we have that $$ \frac dq- \frac d{d+1}+ \ln\left(\frac q{d+1}\right)\leq \frac d{2d+2}- \frac d{d+1}+ \ln\left(\frac q{d+1}\right)< \ln\left(\frac q{d+1}\right).$$
 Now, let us prove that
 \begin{equation}\label{eq:new1}
 \frac dq- \frac d{d+1}+ \ln\left(\frac q{d+1}\right)\geq c \ln\left(\frac q{d+1}\right),
 %>\frac dq- 1+ \ln\left(\frac q{d+1}\right),
 \end{equation} for any positive constant $c\leq \frac 16$. 
 Notice that   inequality (\ref{eq:new1}) holds if and only if
\begin{equation}\label{eq:new2}
1-c \geq\frac {\frac d{d+1}-\frac dq}{ \ln\left(\frac q{d+1}\right)}.\end{equation}

Since $q\geq 2d+2$, the right-hand side of inequality (\ref{eq:new2}) is smaller than
\begin{eqnarray*}
\frac {\frac{q-d-1}q}{ \ln\left(\frac q{d+1}\right)}&=& \frac{\frac{q-d-1}q}{-\ln\left(1-\frac {q-d-1}{q}\right)}
= \frac{1}{-\ln\left(1-\frac {q-d-1}{q}\right)^\frac q{q-d-1}}
\leq \frac{1}{2\ln2}, \label{eq:new3}
\end{eqnarray*}
where the last inequality follows from setting $f=\frac q{q-d-1}$ and observing that $-f\ln (1-1/f)$ decreases with $f$. Since  $q\geq 2d+2$ implies
$f\leq 2$, it holds $-f\ln (1-1/f)\geq 2\ln2$. Therefore, one has that inequality (\ref{eq:new2})  holds for any $c$ such that $1-c\geq  \frac 1{2\ln 2}$.
Since $1- \frac 1{2\ln 2}\geq 1- \frac 1{1.2}=\frac 16$, it follows that inequality (\ref{eq:new1}) holds for any $c\leq \frac 16$.
Therefore,
\begin{equation}
1-H_q(\delta)=\Theta\left(\frac 1{(d+1)\ln q} \ln\left(\frac q{d+1}\right)\right).
\end{equation}

\noindent
It follows that
\begin{equation}\label{eq:prLC3}
\log_q n=k= m(1-H_q(\delta))
= \Theta\left(\frac m{(d+1)\ln q}\ln \left(\frac q{d+1}\right) \right).
\end{equation}
By setting $s=dm$ and $t=mq$ in  (\ref{eq:prLC3}), we get
\begin{equation}\label{eq:prLC4}
\ln n=\log_q n  \ln q=\Theta\left(\frac s{d(d+1)} \ln\left(\frac {td}{s(d+1)}\right) \right)=\Theta\left(\frac s{d^2}\ln \left(\frac {t}{s}\right) \right). 
\end{equation}
The maximum number  $\y_\A(d,n,t)$ of positive responses admitted by the algorithm is equal to the maximum number of elements contained in the union of $d$ members of the family.  Since $s$ is an upper bound on the size of the union of any $d$ members of the family,  one has that $\y_\A(d,n,t)\leq s$.
By Theorem \ref{thm:pr_reducition},  any two members of the family intersect in at most $m-\delta m$ elements. Hence,  Lemma \ref{lemma2} implies  $|\bigcup_{i=1}^d F_{j_i}|\geq md-\frac12(m-\delta m)d(d-1)=s-s(d-1)/(2d+2)\geq s/2$, for any $d$ members $F_{j_1},\ldots,F_{j_d}$ of the family. Therefore, it holds   $s/2\leq \y_\A(d,n,t)\leq s$,  from which
 the first bound for $\y_\A(d,n,t)< \frac{(t+1)d}{2(d+1)}$ in the statement of the theorem follows. 
 
 In order to obtain the bound expressed only in terms of $d$, $t$ and $n$,  we apply recursively the first  bound  to limit  $\y_\A(d,n,t)$ in   its expression, thus obtaining
\begin{equation}\label{eq:upP}
\y_\A(d,n,t)=\Theta
\Bigg( \frac{d^2\ln n}{  
\ln \Big(
\frac t{d^2\ln n}\ln\left(\frac t {\y_\A(d,n,t)}\right)\Big)}\Bigg).
\end{equation}
Since $\y_\A(d,n,t)\leq \frac{(t+1)d}{2(d+1)}$,  we have that the right-hand side of (\ref{eq:upP}) is $O\left( {d^2\ln n
\over  
\log \left(
\frac{t}{d^2\log n}\right)}\right)$, thus obtaining the second bound in the statement of the theorem.

 The time needed to construct the family is $\Theta(q^km)=\Theta\left(n\frac{\y_\A(d,n,t)}d\right)$. By applying the bound  $\y_\A(d,n,t)=\Theta\left({d^2\ln n\over{\ln(\frac t {\y_\A(d,n,t)}})} \right)$, we obtain  $\Theta\left(n\frac{\y_\A(d,n,t)}d\right)=\Theta\left({dn\ln n\over{\ln(\frac t {\y_\A(d,n,t)})} }\right)$,
whereas by applying the right-hand side bound $\y_\A(d,n,t)=O\left( {d^2\ln n
\over  
\ln \left(
\frac{t}{d^2\ln n}\right)}\right)$, we obtain $\Theta\left(n\frac{\y_\A(d,n,t)}d\right)=O\left({dn\ln n\over{\ln(\frac t {d^2\ln n}} )}\right)$. 
 \qed
 %We remark that the constant hidden in the asymptotic upper bound implied by Theorem \ref{thm:explicit_construction} is smaller than 7.
 \section{Optimal two-stage group testing}\label{sec:two_stage}
 We consider trivial two-stage algorithms, i.e., algorithms that consist of two  non-adaptive stages, with the first stage
 performing parallel tests on pools of elements, and the second stage performing individual tests on certain selected elements.
  More precisely, in the first stage a non-adaptive 
 group testing algorithm is used to determine a ``small" number of potential defective elements, i.e., a subset of elements that contains all defectives;
 in the second stage the subset of elements selected by the first stage are individually tested so as to find those that are really defective.
In this section we give a trivial two-stage algorithm that admits the same maximum number of ``yes" responses as the optimal adaptive algorithm, thus showing that by allowing just  a little adaptiveness, one can an achieve the same performance as the best adaptive algorithms.

In the following, given a trivial two-stage algorithm $\A$ that  finds up to $d$ defective elements in an input set of size $n$ by at most $t$ tests, we denote by $\yb_\A(n,d,t)$ the maximum  number of positive responses that may occur during the search process performed by $\A$,
 where the maximum is taken over all possible  subsets of up to $d$ defectives. Moreover, we denote by $\yb(n,d,t)$ the minimum value of $\yb_\A(n,d,t)$  over all trivial two-stage strategies $\A$ that  find up to $d$ defective elements in an input set of size $n$ by at most $t$ tests.

As observed in Section \ref{sec:new_var}, a $\cup_{\leq s}$ $(p,d)$-cover free family can be used to design a non-adaptive algorithm
that selects a subset of up to $p+d-1$ elements containing all defective elements and admits at most $s$ ``yes" responses. Therefore, such an algorithm can be employed in the first stage of a trivial two-stage algorithm to select the elements that will undergo individual tests during the second stage.  Notice that  the total number of positive responses admitted by the two-stage algorithm is at most $s+d$, since at most $d$ individual probes yield a positive response in the second stage.

The following theorem follows from the above discussion.
\begin{theorem}\label{thm:gen_two-stage}
Let $t$, $n$, $d$, $p$ be positive integers with $t\geq   d+p$ and $d+p\leq n\leq n_{cf}(p,d,\cup_{\leq s},t-d-p+1)$. There exists a two-stage group testing strategy $\A$ for which
$$\yb_\A(d,n,t) \leq s+d.
  $$
\end{theorem}

The following theorem is a consequence of Theorem \ref{thm:gen_two-stage} and Theorem \ref{thm:p_d_cf_upp}.
\begin{theorem}\label{thm:two-stage}
Let $t$, $n$, $d$, $p$ be positive integers  with $t\geq d+p$ and $n\geq d+p$. There exists a two-stage group testing strategy $\A$ for which $\yb_\A(d,n,t)$ is at most
$$ \cases{ {d(d+p)\over p}\log\left(\frac {en}{d+p}\right)+d \log \left({e(d+p)\over p}\right)+\frac dp+d &if $\yb_\A(d,n,t) >(t+d-p+1)/2$, \cr\cr
{{d(d+p) \over  p}\log\left({en\over d+p}\right)+d\log\left({e(d+p)\over p}\right)+\frac dp\over \log \left(\frac{e(t-d-p+1)}{\yb_A(d,n,t)-d}\right)}+d
 \leq {{d(d+p) \over  p}\log\left({en\over d+p}\right)+d\log\left({e(d+p)\over p}\right)+\frac dp\over \log \chi}+d &if $\yb_\A(d,n,t) \leq (t+d-p+1)/2$,\\}
  $$
  where $\chi={e(t-d-p+1)\log (2e)\over {d(d+p)\over p}\left(\log \left({en\over d+p}\right)+\log(e\sqrt2)\right)}$.
\end{theorem}
\proof
The two-stage algorithm consists in a first stage in which the pools corresponding to the rows of  a $\cup_{\leq s}$ $(p,d)$-cover free family 
are tested in parallel, and in a second stage that performs individual probes on the up to $d+p-1$ elements selected by the first stage. 
The bound in the statement of the  theorem 
follows from the lower  bound of Theorem  \ref{thm:p_d_cf_upp}
on the maximum size of a $\cup_{\leq s}$ $(p,d)$-cover free family  on the ground set $[t-d-p+1]$. 
The lower  bound of Theorem  \ref{thm:p_d_cf_upp} implies that the number of positive responses in the first stage is
$$s\leq\cases{  {d(d+p)\over p}\log\left(\frac {en}{d+p}\right)+d \log \left({e(d+p)\over p}\right)+\frac dp&if $s>(t-d-p+1)/2$, \cr\cr
{{d(d+p) \over  p}\left(\log\left({en\over d+p}\right)+\frac p{ d+p}\log\left({e(d+p)\over p}\right)+\frac 1{d+p}\right)\over \log \left(\frac{e(t-d-p+1)}{s}\right)}&if $s\leq(t-d-p+1)/2$.\\}$$

%Observe that, the above two bounds are both 
%$$O\left( {{(d+p)^2}\log\left({n\over d+p}\right) \over  p\log \left(\frac{e(t-d-p)}{s-d}\right)}\right).$$ This is obvious for the second bound and it is a consequence of the fact that $s-d=\Theta(t-d-p)$ as far as it concerns  the first bound.
Since up  to $d$ individual probes yield a positive response in the second stage, we set $s=\yb_\A(d,n,t)-d$ so that  the algorithm is guaranteed to receive no more that  $\yb_\A(d,n,t)$ ``yes" responses in total. 
By setting $s=\yb_\A(d,n,t)-d$  in the above bounds, we get the  bound for $\yb_\A(d,n,t) >(t+d-p+1)/2$ in the statement of the theorem and the first of the two bounds stated  for $\yb_\A(d,n,t) \leq (t+d-p+1)/2$ . In order to obtain the second bound for $\yb_\A(d,n,t) \leq (t+d-p+1)/2$, we first observe that
$\frac p{ d+p}\log\left({e(d+p)\over p}\right)+\frac 1{d+p}$ decreases with $d$ and consequently is smaller than 
$\frac p{ 1+p}\log\left({e(1+p)\over p}\right)+\frac 1{1+p}\leq \log (e\sqrt 2)$.
Therefore, we have that
$$
\yb_A(d,n,t)\leq {{d(d+p) \over  p}\left(\log\left({en\over d+p}\right)+\log (e\sqrt 2)\right)\over \log \left(\frac{e(t-d-p+1)}{\yb_A(d,n,t)-d}\right)}+d.
$$
Then, we bound $\yb_\A(d,n,t)$ by $(t+d-p+1)/2$ in the above  upper bound, thus obtaining 
\begin{equation}\label{eq:4_8_2}
\yb_A(d,n,t)\leq {{d(d+p) \over  p}\left(\log\left({en\over d+p}\right)+\log (e\sqrt 2)\right)\over \log (2e)}+d.
\end{equation}
We exploit upper bound (\ref{eq:4_8_2}) to limit from above 
$\yb_\A(d,n,t)$ in the first of the two bounds stated for  $\yb_\A(d,n,t) \leq (t+d-p+1)/2$, thus getting the second bound for $\yb_\A(d,n,t) \leq (t+d-p+1)/2$ in the statement of
the theorem. 
\qed

By setting $p=d$ in the bound of Theorem \ref{thm:two-stage}, we obtain the following corollary that states the existence of a
trivial
two-stage algorithm which asymptotically attains the same  bound of the optimal adaptive algorithm. 
\begin{corollary}\label{cor:two-stage}
Let $t$, $n$, $d$ be positive integers  with $t\geq 2d$ and $n\geq 2d$. There exists a two-stage group testing strategy $\A$ for which
$$\yb_\A(d,n,t) \leq \cases{ {2d}\log\left(\frac {en}{2d}\right)+d \log \left({2e}\right)+d+1 &if $\yb_\A(d,n,t) >t/2$, \cr\cr
{2d\log\left({en\over 2d}\right)+d\log(2e)+1\over \log \left(\frac{e(t-2d+1)}{\yb_A(d,n,t)-d}\right)}+d
 \leq {{2d}\log\left({en\over 2d}\right)+d\log({2e})+1\over \log \chi' }+d &if $\yb_\A(d,n,t) \leq t/2$,\\}
  $$
  where $\chi'={e(t-2d+1)\log (2e)\over {2d}\left(\log \left({en\over 2 d}\right)+\log(e\sqrt 2)\right)}$.
%$$\yb_\A(d,n,t) = O\left( {d\log\left({n\over d}\right) \over  \log \left(\frac{e(t-2d)}{\yb_A(d,n,t)-d}\right)}\right)=
%O\left( {d\log\left({n\over d}\right) \over  \log \left(\frac{e(t-2d)}{d\log\left({n\over d}\right)-d}\right)}\right).$$
\end{corollary}

\end{document}